\documentclass[10pt]{article}
\usepackage{amsmath}
\usepackage{amssymb}
\usepackage[cmtip,arrow]{xy}
\usepackage{pb-diagram,pb-xy}

\begin{document}

\title{\textbf{Integrable systems on semidirect product Lie groups}\\
\smallskip\ }
\author{\textbf{S. Capriotti$^{\dag }$ \& H. Montani$^{\ddag }${\thanks{%
e-mail: \textit{\ hmontani@uaco.unpa.edu.ar }} } } \\
\dag\ \textit{Departamento de Matem\'{a}tica, Universidad Nacional del
Sur,\smallskip }\ \\
\ ~\textit{Av. Alem 1253, 8000} - \textit{Bah\'{\i}a Blanca, Buenos Aires,
Argentina. }\\
\\
\ddag\ \textit{Departamento de Ciencias Exactas y Naturales, }\\
\textit{Unidad Acad\'{e}mica Caleta Olivia,}\\
\textit{Universidad Nacional de la Patagonia Austral.}\\
\textit{9011 - Caleta Olivia, Santa Cruz, Argentina}\\
}
\maketitle

\begin{abstract}
We study integrable systems on the semidirect product of a Lie group and its
Lie algebra as the representation space of the adjoint action. Regarding the
tangent bundle of a Lie group as phase space endowed with this semidirect
product Lie group structure, we construct a class of symplectic submanifolds
equipped with a Dirac bracket on which integrable systems (in the
Adler-Kostant-Symes sense) are naturally built through collective dynamics.
In doing so, we address other issues as factorization, Poisson-Lie
structures and dressing actions. We show that the procedure becomes recursive
for some particular Hamilton functions, giving rise to a tower of nested
integrable systems.
\end{abstract}

\tableofcontents

\newpage

\newpage

\section{Introduction}

Integrable hamiltonian systems find a natural setting in the realm of Lie
algebras and Lie groups where their equations of motion are realized as Lax
pairs. It becomes an standard framework for many integrable systems after
the seminal work by Arnold \cite{Arnold} encoding in this framework the
equations of motion associated with the rigid body and incompressible fluid.
This setting fits perfectly for systems which are strongly symmetric, in
such a way that the configuration space can be identified with the symmetry
group.

Those systems with a Lie group as configuration space have the cotangent
bundle of this Lie group as phase space, which in turn can be identified
with the Cartesian product of the group itself and the dual of its Lie
algebra. They are symplectic manifolds, meaning that the Poisson brackets
are nondegenerate, enjoying many nice properties related to the symmetry
issues \cite{Abr-Mars},\cite{Mars-Ratiu}. The restriction of this bracket to
functions on the dual of the Lie algebra produces the Lie-Poisson bracket
whose symplectic leaves coincides with coadjoint orbits. When the Lie
algebra is supplied with a nondegenerate Ad-invariant bilinear form, one can
translate the Poisson structure to the Lie algebra and the equations of
motion turns into the Lax pair form.

It is surprising that when symmetries are broken one may still find a Lie
groups-Lie algebras setting encoding these kind of systems by considering
semidirect products of Lie groups \cite{Vino-Kuper}. This is the case when
potential energy is included as, for instance, in the rigid body system and
its N-dimensional analogues \cite{Ratiu-vanMoerb},\cite{Ratiu 0},\cite{Ratiu
1}. The general setting is that reduction of the cotangent bundle of a Lie
group by the action of some Lie subgroup makes semidirect products arise. A
deep understanding of this connection began to appear in ref. \cite%
{Guillemin-Sternberg}, and it was fully clarified in ref. \cite{MWR}, where
the heavy top and compressible flow are presented as motivating examples.
There are many other dynamical systems falling in this scheme as the
Kirchhoff equation for the motion of a rigid body in an ideal incompressible
fluid moving under a potential and at rest at infinity, and the Leggett
equation for the magnetic moment in the low temperature phases of $^{3}He$ 
\cite{Novikov},\cite{GNM}, where other issues of this kind of systems are
analyzed. Since then, further developments and applications has been carried
out widening the involvement of semidirect product Lie groups in dynamical
systems \cite{HMR 1},\cite{MMPR},\cite{CHMR},\cite{Holm-Tronci}.

In this work we investigate integrable systems on semidirect product Lie
groups from the point of view of the Adler-Kostant-Symes (AKS) \cite{Adler},%
\cite{Kostant},\cite{Symes} and Reyman and Semenov-Tian-Shansky \cite{RSTS 0}%
,\cite{RSTS 1} theory. In this context, non trivial integrable systems are
generated by a restriction procedure: given a Lie algebra $\mathfrak{g}=%
\mathfrak{g}_{+}\oplus \mathfrak{g}_{-}$, with $\mathfrak{g}_{+},\mathfrak{g}%
_{-}$ Lie subalgebras of $\mathfrak{g}$, the restriction of the set of $Ad$%
-invariant functions on a coadjoint orbit of $\mathfrak{g}$ to the coadjoint
orbit of one of its components, for instance $\mathfrak{g}_{+}$, gives rise
to a non trivial set of Poisson commuting functions, so that the
Arnold-Liouville theorem holds. These ideas are extended to cotangent
bundles of the Lie groups $G,G_{+},G_{-}$ associated with $\mathfrak{g},%
\mathfrak{g}_{+},\mathfrak{g}_{-}$, and the solution of the equations of
motion on the cotangent bundle of the factors $G_{+}$, for instance, arises
as the $G_{+}$-factor of an exponential curve in $G$.

As the general framework, we consider the semidirect product of a Lie group
with the underlying vector space of its Lie algebra regarded as the
representation space of the adjoint action and, for dynamical issues, the
Dirac scheme as presented in ref. \cite{CapMon JMP} focusing our attention
on a family of symplectic submanifolds in $G\times \mathfrak{g}$ defined as
the level set of the \emph{projection map }$\Psi :G\times \mathfrak{g}%
\longrightarrow G_{-}\times \mathfrak{g}_{-}$.

We study hamiltonian systems on the tangent bundle of a semidirect product
Lie group and work out the structure of the Hamilton equations of motion by
writing them out in terms of its original components. In particular,
collective dynamics and factorization will connect our construction with
integrable systems. Because the factorization arises from a Poisson-Lie
group, dressing vector fields are strongly involved in the dynamics, making
relevant to analyze the induced Poisson-Lie structure on the semidirect
products and the associated dressing actions. The construction produces a
set of equations of motion resembling those of the heavy top just by
substituting the adjoint action by the dressing one, and they can be solved
by factorization following the AKS ideas. Also, for some particular
collective hamiltonians, the construction can be promoted to a recursive
procedure on iterated semidirect products, leading to a tower of integrable
systems. Some issues of this developments are showed in an explicit example
built on $SL(2,\mathbb{C)}$ and its Iwasawa decomposition.

The work is ordered as follows: in Section II we fix the algebraic tools of
the problem, in Section III we describe the involved phase spaces, with the
corresponding Dirac brackets, working out symmetries, the additional
Poisson-Lie structure and the associated dressing actions. In Section IV we
focus on the dynamical systems, collective dynamics and integrability by
factorization describing also the nested equation of motion inherited from
the nested semidirect product structure. In Section V, we build up a tower
of integrable system for a particular collective hamiltonian, and finally,
in Section VI we present an example on $SL(2,\mathbb{C)}$ showing the
explicit solution obtained by factorization.

\section{Semidirect products with the adjoint representation}

Let $H_{1}$ be a Lie group and $\mathfrak{h}_{1}$ its Lie algebra, which we
assume equipped with a nondegenerate symmetric bilinear form 
\begin{equation*}
\mathsf{k}_{1}:\mathfrak{h}_{1}\otimes \mathfrak{h}_{1}\longrightarrow 
\mathbb{R}
\end{equation*}%
Associated to them, we consider the semidirect product $H_{2}=H_{1}\circledS 
\mathfrak{h}_{1}$, where the vector space $\mathfrak{h}_{1}$, with the
trivial Lie algebra structure, is regarded as the representation space for
the \emph{adjoint action} of $H_{1}$ in the right action structure of
semidirect product, so the Lie group structure for $\left( a,X\right)
,\left( b,Y\right) \in H_{2}$ is 
\begin{equation}
\left( a,X\right) \bullet \left( b,Y\right) =\left( ab,\mathrm{Ad}%
_{b^{-1}}^{1}X+Y\right)  \label{Hxh-1}
\end{equation}%
where $\mathrm{Ad}^{1}$ stands for the adjoint action of $H_{1}$ on its Lie
algebra $\mathfrak{h}_{1}$.

In this setting, the right and left translation applied on $\left(
v,Z\right) \in T_{(a,X)}H_{2}$ reads:%
\begin{equation*}
\left\{ 
\begin{array}{l}
(R_{\left( b,Y\right) }^{\bullet })_{\ast \left( a,X\right) }\left(
v,Z\right) =\left( \left( R_{b}\right) _{\ast }v,\mathrm{Ad}%
_{b^{-1}}^{1}Z\right) _{(ab,\mathrm{Ad}_{b^{-1}}^{1}X+Y)} \\ 
\\ 
(L_{\left( b,Y\right) }^{\bullet })_{\ast \left( a,X\right) }\left(
v,Z\right) =\left( (L_{b})_{\ast }v,\mathrm{Ad}_{a^{-1}}^{1}\left[
Y,(R_{a^{-1}})_{\ast }v\right] +Z\right) _{(ba,\mathrm{Ad}_{a^{-1}}^{1}Y+X)}%
\end{array}%
\right. \,
\end{equation*}%
Let $\left( X,U\right) \in \mathfrak{h}_{2}=\mathfrak{h}_{1}\oplus \mathfrak{%
h}_{1}$. So, we have that the right and left invariant vector fields on $%
H_{2}=H_{1}\circledS \mathfrak{h}_{1}$ at a point $\left( a,V\right) $ are
defined as 
\begin{equation*}
\left\{ 
\begin{array}{l}
\left. \left( X,U\right) ^{R\bullet }\right\vert _{\left( a,V\right)
}:=(R_{\left( a,V\right) }^{\bullet })_{\ast \left( e,0\right) }\left(
X,U\right) =\left( Xa,\mathrm{Ad}_{a^{-1}}^{1}U\right) _{(a,V)} \\ 
\\ 
\left. \left( X,U\right) ^{L\bullet }\right\vert _{\left( a,V\right)
}:=(L_{\left( a,V\right) }^{\bullet })_{\ast \left( e,0\right) }\left(
X,U\right) =\left( aX,\left[ V,X\right] +U\right) _{(a,V)}%
\end{array}%
\right.
\end{equation*}%
The exponential map $\mathrm{Exp}^{\bullet }:\mathfrak{h}_{2}\mathbf{%
\longrightarrow }H_{2}$ is 
\begin{equation}
\mathrm{Exp}^{\bullet }\left( t\left( X,Y\right) \right) =\left(
e^{tX},-\left( \sum_{n=1}^{\infty }\frac{\left( -1\right) ^{n}}{n!}%
t^{n}\left( \mathrm{ad}_{X}^{1}\right) ^{n-1}\right) Y\right)  \label{Hxh-4}
\end{equation}%
and the adjoint action of $H_{2}$ on $\mathfrak{h}_{2}$ is given by%
\begin{equation}
\mathrm{Ad}_{\left( b,Z\right) }^{2}\left( X,Y\right) =\left( \mathrm{Ad}%
_{b}^{1}X\,,\mathrm{Ad}_{b}^{1}\left( \left[ Z,X\right] _{1}+Y\right) \right)
\label{Hxh-5}
\end{equation}%
The Lie algebra structure on $\mathfrak{h}_{2}=\mathfrak{h}_{1}\circledS 
\mathfrak{h}_{1}$ is%
\begin{equation}
\left[ \left( Y,V\right) ,\left( X,Z\right) \right] _{2}=\mathrm{ad}_{\left(
Y,V\right) }^{2}\left( X,Z\right) =\left( \left[ Y,X\right] _{1},\left[ Y,Z%
\right] _{1}+\left[ V,X\,\right] _{1}\right)  \label{Hxh-6}
\end{equation}%
underlying the Lie group semidirect product structure of $%
H_{2}=H_{1}\circledS \mathfrak{h}_{1}$. The coadjoint action becomes in%
\begin{equation*}
\mathrm{ad}_{\left( Y,V\right) }^{2\ast }\left( \eta ,\xi \right) =\left( 
\mathrm{ad}_{Y}^{1\ast }\eta +\mathrm{ad}_{V}^{1\ast }\xi ,\mathrm{ad}%
_{Y}^{1\ast }\xi \right)
\end{equation*}

\subsection{Bilinear forms and factorization}

We now consider the Lie algebra $\mathfrak{h}_{1}$ equipped with the
symmetric nondegenerate $\mathrm{Ad}^{1}$-invariant bilinear form $\mathsf{k}%
_{1}:\mathfrak{h}_{1}\otimes \mathfrak{h}_{1}\longrightarrow \mathbb{R}$,
and we define on $\mathfrak{h}_{2}$ the nondegenerate symmetric bilinear
form $\mathsf{k}_{2}:\mathfrak{h}_{2}\otimes \mathfrak{h}_{2}\longrightarrow 
\mathbb{R}$ as 
\begin{equation}
\mathsf{k}_{2}\left( \left( X,U\right) ,\left( Y,V\right) \right) :=\mathsf{k%
}_{1}\left( X,V\right) +\mathsf{k}_{1}\left( Y,U\right)  \label{Hxh-10}
\end{equation}%
which is\textit{\ }also $\mathrm{Ad}^{2}$-invariant.

Let us assume that there exists a factorization $H_{1}=H_{1}^{+}H_{1}^{-}$
where $H_{1}^{+}$ and $H_{1}^{-}$ are Lie subgroups of $H_{1}$, implying
that every $h_{1}\in H_{1}$ can be written as 
\begin{equation}
h_{1}=h_{1}^{+}h_{1}^{-}  \label{factor H1}
\end{equation}%
for some $h_{1}^{+}\in H_{1}^{+}$ and $h_{1}^{-}\in H_{1}^{-}$.
Consequently, $\mathfrak{h}_{1}=\mathfrak{h}_{1}^{+}\oplus \mathfrak{h}%
_{1}^{-}$ with $\mathfrak{h}_{1}^{+}$ and $\mathfrak{h}_{1}^{-}$ being Lie
subalgebras of $\mathfrak{h}_{1}$. The factorization at this level can be
seen in the crossed Lie bracket\textit{\ } 
\begin{equation}
\left[ X_{1}^{-},X_{1}^{+}\right] =\left( X_{1}^{+}\right)
^{X_{1}^{-}}+\left( X_{1}^{-}\right) ^{X_{1}^{+}}  \label{dlg-2a}
\end{equation}%
where\textit{\ }$X_{1}^{+},\left( X_{1}^{+}\right) ^{X_{1}^{-}}\in \mathfrak{%
h}_{1}^{+}$ and\textit{\ }$X_{1}^{-},\left( X_{1}^{-}\right) ^{X_{1}^{+}}\in 
\mathfrak{h}_{1}^{-}$. Also we assume that $\mathfrak{h}_{1}^{+},\mathfrak{h}%
_{1}^{-}$ are \emph{isotropic subspaces} in relation with $\mathsf{k}_{1}:%
\mathfrak{h}_{1}\otimes \mathfrak{h}_{1}\longrightarrow \mathbb{R}$, such
that $\left( \mathfrak{h}_{1},\mathfrak{h}_{1}^{+},\mathfrak{h}%
_{1}^{-}\right) $ compose a \emph{Manin triple}. In this way, each Lie
algebra is also \emph{Lie bialgebra} and the associated groups are \emph{%
Poisson-Lie groups}.

The factorization of $H_{1}$ induces the factorization $H_{2}=H_{2}^{+}%
\bullet H_{2}^{-}$, where $H_{2}^{\pm }=H_{1}^{\pm }\circledS \mathfrak{h}%
_{1}^{\pm }$, such that any element $\left( h_{1},X_{1}\right) \in H_{2}$
can be written as%
\begin{equation*}
\left( h_{1},Y_{1}\right) =\left( h_{1}^{+},X_{1}^{+}\right) \bullet \left(
h_{1}^{-},X_{1}^{-}\right) =\left( h_{1}^{+}h_{1}^{-},\mathrm{Ad}%
_{h_{-}^{-1}}^{1}X_{1}^{+}+X_{1}^{-}\right)
\end{equation*}%
with%
\begin{equation}
\left\{ 
\begin{array}{l}
h_{1}=h_{1}^{+}h_{1}^{-} \\ 
\\ 
X_{1}^{+}=\Pi _{+}\left( \mathrm{Ad}_{h_{1}^{-}}^{1}Y_{1}\right) \\ 
\\ 
X_{1}^{-}=\mathrm{Ad}_{h_{-}^{-1}}^{1}\Pi _{-}\mathrm{Ad}%
_{h_{1}^{-}}^{1}Y_{1}%
\end{array}%
\right.  \label{factorization semidirect 2}
\end{equation}

\begin{description}
\item[Remark:] \textit{Observe the above factorization is fully determined
by factorization in the previous step: let} $h_{2}=\left( h_{1},Y_{1}\right)
\in H_{2}$, $h_{1}\in H_{1}$ \textit{and} $Y_{1}\in \mathfrak{h}_{1}$\textit{%
, then}%
\begin{equation*}
h_{2}=\left( h_{1}^{+},\Pi _{+}\mathrm{Ad}_{h_{1}^{-}}^{1}Y_{1}\right) \cdot
\left( h_{1}^{-},\mathrm{Ad}_{\left( h_{1}^{-}\right) ^{-1}}^{1}\Pi _{-}%
\mathrm{Ad}_{h_{1}^{-}}^{1}Y_{1}\right)
\end{equation*}%
\textit{where} $h_{1}^{+}$ \textit{and} $h_{1}^{-}$ \textit{are such that }$%
h_{1}=h_{1}^{+}h_{1}^{-}$.
\end{description}

The Lie algebra $\mathfrak{h}_{2}$ decomposes in a direct sum as $\mathfrak{h%
}_{2}=\mathfrak{h}_{2}^{+}\oplus \mathfrak{h}_{2}^{-}$, with $\mathfrak{h}%
_{2}^{\pm }=\mathfrak{h}_{1}^{\pm }\circledS \mathfrak{h}_{1}^{\pm }$ being
Lie subalgebras of $\mathfrak{h}_{2}$ such that for $\left( X,V\right) \in 
\mathfrak{h}_{2}$, 
\begin{equation*}
\left( X,V\right) =\left( X_{1}^{+},V_{1}^{+}\right) +\left(
X_{1}^{-},V_{1}^{-}\right)
\end{equation*}%
Moreover, $\mathfrak{h}_{2}^{+}$ and $\mathfrak{h}_{2}^{-}$ are \emph{%
isotropic subspaces}\textit{\ }in relation with $\mathsf{k}_{2}$ implying
the bijections $\gamma \left( \mathfrak{h}_{2}^{\pm }\right) =\left( 
\mathfrak{h}_{2}^{\mp }\right) ^{\ast }$ and $\left( \mathfrak{h}_{2}^{\pm
}\right) ^{\bot }=\mathfrak{h}_{2}^{\pm }$. For the sake of simplicity, we
shall use $\gamma $ to generically denote the bijections induced by the
bilinear forms $\mathsf{k}_{i}$, for any $i$, no confusion would arise from
this ambiguity since it becomes clear which map must be used from the labels
of its arguments.

Let us quote some useful relations holding for every Lie bialgebra and
Poisson-Lie group. They can be obtained from the interplay of the relation $%
\left( \ref{dlg-2a}\right) $ and the bilinear forms $\mathsf{k}_{i}$: 
\begin{equation}
\begin{array}{ccc}
X_{+}^{Y_{-}}=-\gamma ^{-1}\left( ad_{Y_{-}}^{\ast }\gamma \left(
X_{+}\right) \right) & , & X_{+}^{h_{-}}=\gamma ^{-1}\left(
Ad_{h_{-}^{-1}}^{\ast }\gamma \left( X_{+}\right) \right)%
\end{array}
\label{R1}
\end{equation}%
and 
\begin{equation}
\mathrm{Ad}_{h_{+}^{-1}}X_{-}=h_{+}^{-1}h_{+}^{X_{-}}+\gamma ^{-1}\left(
Ad_{h_{+}}^{\ast }\gamma \left( X_{-}\right) \right)  \label{R2}
\end{equation}%
We remark that $\mathrm{Ad}$ denotes the adjoint action of the groups $%
H=H^{+}H^{-}$, while $Ad$ denotes the corresponding adjoint actions of the
factors $H^{+}$ or $H^{-}$.

\subsection{Poisson-Lie structure on $H_{2}\label{PL structure}$}

The Lie algebras $\mathfrak{h}_{i}$, $\mathfrak{h}_{i}^{+}$ and $\mathfrak{h}%
_{i}^{-}$ besides the bilinear form $\mathsf{k}_{i}:\mathfrak{h}_{i}\otimes 
\mathfrak{h}_{i}\longrightarrow \mathbb{R}$, for $i=1,2$, constitute \emph{%
Manin triples }$\left( \mathfrak{h}_{i},\mathfrak{h}_{i}^{+},\mathfrak{h}%
_{i}^{-}\right) $ so, each Lie algebra is indeed a Lie bialgebra, and the
associated Lie groups are Poisson-Lie groups. For instance, on $H_{2}^{+}$
the Poisson-Lie structure $\pi _{2}^{+}$ is defined as, see ref. \cite{Lu-We}%
, 
\begin{eqnarray}
&&\left\langle \gamma \left( X_{1}^{-},V_{1}^{-}\right) \left(
h_{1}^{+},Z_{1}^{+}\right) ^{-1}\otimes \gamma \left(
Y_{1}^{-},W_{1}^{-}\right) \left( h_{1}^{+},Z_{1}^{+}\right) ^{-1},\pi
_{2}^{+}\left( h_{1}^{+},Z_{1}^{+}\right) \right\rangle  \notag \\
&=&\mathsf{k}_{2}\left( \Pi _{-}\mathrm{Ad}_{\left(
h_{1}^{+},Z_{1}^{+}\right) ^{-1}}^{2}\left( X_{1}^{-},V_{1}^{-}\right) ,\Pi
_{+}\mathrm{Ad}_{\left( h_{1}^{+},Z_{1}^{+}\right) ^{-1}}^{2}\left(
Y_{1}^{-},W_{1}^{-}\right) \right)  \label{Poisson Lie struc 0}
\end{eqnarray}%
Having in mind the expression $\left( \ref{Hxh-5}\right) $, and the
Poisson-Lie structure in $H_{1}^{+}$%
\begin{eqnarray*}
&&\left\langle \gamma \left( X_{1}^{-}\right) \left( h_{1}^{+}\right)
^{-1}\otimes \gamma \left( Y_{1}^{-}\right) \left( h_{1}^{+}\right)
^{-1},\pi _{1}^{+}\left( h_{1}^{+}\right) \right\rangle \\
&=&\mathsf{k}_{1}\left( \Pi _{-}\mathrm{Ad}_{\left( h_{1}^{+}\right)
^{-1}}^{1}X_{1}^{-},\Pi _{+}\mathrm{Ad}_{\left( h_{1}^{+}\right)
^{-1}}^{1}Y_{1}^{-}\right)
\end{eqnarray*}%
we may write 
\begin{equation}
\begin{array}{l}
\mathsf{k}_{2}\left( \left( X_{1}^{-},V_{1}^{-}\right) \otimes \left(
Y_{1}^{-},W_{1}^{-}\right) ,\left( R_{\left( h_{1}^{+},Z_{1}^{+}\right)
^{-1}}\right) _{\ast }^{\otimes 2}\pi _{2}^{+}\left(
h_{1}^{+},Z_{1}^{+}\right) \right) \\ 
\\ 
~~~=\mathsf{k}_{1}\left( X_{1}^{-}\otimes W_{1}^{-}+V_{1}^{-}\otimes
Y_{1}^{-},\left( R_{h_{+}^{-1}}\right) _{\ast }^{\otimes 2}\pi
_{1}^{+}\left( h_{1}^{+}\right) \right) \\ 
\\ 
~~~~~~+\mathsf{k}_{1}\left( X_{1}^{-}\otimes Y_{1}^{-},\left( \mathrm{ad}_{%
\mathrm{Ad}_{h_{1}^{+}}^{1}Z_{1}^{+}}^{1}\right) ^{\otimes 2}\left(
R_{h_{+}^{-1}}\right) _{\ast }^{\otimes 2}\pi _{1}^{+}\left(
h_{1}^{+}\right) \right)%
\end{array}
\label{Poisson Lie struc 2}
\end{equation}%
where $\mathsf{k}_{1}\left( A\otimes B,C\otimes D\right) =\mathsf{k}%
_{1}\left( A,C\right) \mathsf{k}_{1}\left( B,D\right) $. In terms of the
components in $\mathfrak{h}_{1}^{+}\oplus \mathfrak{h}_{1}^{-}$ it turns
into 
\begin{equation}
\begin{array}{l}
\mathsf{k}_{2}\left( \left( X_{1}^{-},V_{1}^{-}\right) \otimes \left(
Y_{1}^{-},W_{1}^{-}\right) ,\left( R_{\left( h_{1}^{+},Z_{1}^{+}\right)
^{-1}}\right) _{\ast }^{\otimes 2}\pi _{2}^{+}\left(
h_{1}^{+},Z_{1}^{+}\right) \right) \\ 
\\ 
~~~=\mathsf{k}_{1}\left( X_{1}^{-},\left( h_{1}^{+}\right)
^{W_{1}^{-}}\left( h_{1}^{+}\right) ^{-1}\right) +\mathsf{k}_{1}\left(
V_{1}^{-},\left( h_{1}^{+}\right) ^{Y_{1}^{-}}\left( h_{1}^{+}\right)
^{-1}\right) \\ 
\\ 
~~~~~~+\mathsf{k}_{1}\left( Y_{1}^{-},\left[ \left( h_{1}^{+}\right)
^{X_{1}^{-}}\left( h_{1}^{+}\right) ^{-1},\mathrm{Ad}%
_{h_{1}^{+}}^{1}Z_{1}^{+}\right] _{1}\right) \\ 
\\ 
~~~~~~-\mathsf{k}_{1}\left( X_{1}^{-},\left[ \left( h_{1}^{+}\right)
^{Y_{1}^{-}}\left( h_{1}^{+}\right) ^{-1},\mathrm{Ad}%
_{h_{1}^{+}}^{1}Z_{1}^{+}\right] _{1}\right) \\ 
\\ 
~~~~~~+\mathsf{k}_{1}\left( \left[ X_{1}^{-},Y_{1}^{-}\right] _{1},\mathrm{Ad%
}_{h_{1}^{+}}^{1}Z_{1}^{+}\right)%
\end{array}
\label{Poisson Lie struc 4}
\end{equation}

The Poisson-Lie bracket for a couple of functions $F,G$ on $H_{2}^{+}$ is%
\begin{equation}
\left\{ F,G\right\} _{2}^{+}\left( h_{2}^{+}\right) =\left\langle dF\otimes
dG,\pi _{2}^{+}\left( h_{2}^{+}\right) \right\rangle  \label{PL bracket}
\end{equation}%
In order to take account of the structure of semidirect product of $%
H_{2}^{+} $, we write its differentials as $dF=\left( \mathbf{d}F,\delta
F\right) \in T^{\ast }H_{1}^{+}\oplus \mathfrak{h}_{1}^{+}$. Due to the
relation between $\mathsf{k}_{1}$ and $\mathsf{k}_{2}$, eq. $\left( \ref%
{Hxh-10}\right) $, we have that the bijections $\gamma _{i}:\mathfrak{h}%
_{i}\longrightarrow \left( \mathfrak{h}_{i}\right) ^{\ast }$, $i=1,2$, are
related as 
\begin{equation*}
\gamma _{2}\left( X_{1},X_{2}\right) =\left( \gamma _{1}\left( X_{2}\right)
,\gamma _{1}\left( X_{1}\right) \right)
\end{equation*}%
Also, we need the right translations on the cotangent bundle 
\begin{equation}
R_{\left( b,Y\right) }^{\ast }\left( \alpha ,\beta \right) =\left(
R_{b}^{\ast }\alpha ,\mathrm{Ad}_{b^{-1}}^{1\ast }\beta \right)
\label{R trans on forms}
\end{equation}%
for $\left( \alpha ,\beta \right) \in T_{\left( a,X\right) \cdot \left(
b,Y\right) }^{\ast }H_{2}$. So, comparing $\left( \ref{PL bracket}\right) $
with the expression $\left( \ref{Poisson Lie struc 2}\right) $, we get 
\begin{eqnarray*}
\left\{ dF,dG\right\} _{2}^{-}\left( h_{1}^{+},Z_{1}^{+}\right)
&=&\left\langle \delta F\otimes \mathbf{d}G,\left( \left(
L_{h_{+}^{-1}}\right) _{\ast }\otimes id\right) \pi _{1}^{+}\left(
h_{1}^{+}\right) \right\rangle \\
&&+\left\langle \mathbf{d}F\otimes \delta G,\left( id\otimes \left(
L_{h_{+}^{-1}}\right) _{\ast }\right) \pi _{1}^{+}\left( h_{1}^{+}\right)
\right\rangle \\
&&+\left\langle \delta F\otimes \delta G,\left( \mathrm{ad}%
_{Z_{1}^{+}}^{1}\right) ^{\otimes 2}\left( L_{h_{+}^{-1}}\right) _{\ast
}^{\otimes 2}\pi _{1}^{+}\left( h_{1}^{+}\right) \right\rangle
\end{eqnarray*}

This expression suggest to write the Poisson-Lie bivector in a block matrix
form on $\mathfrak{h}_{1}^{+}\oplus \mathfrak{h}_{1}^{+}$ as 
\begin{equation}
\pi _{2}^{+}\left( h_{1}^{+},Z_{1}^{+}\right) =\left( 
\begin{array}{cc}
0 & \left( id\otimes \left( L_{h_{+}^{-1}}\right) _{\ast }\right) \pi
_{1}^{+}\left( h_{1}^{+}\right) \\ 
\left( \left( L_{h_{+}^{-1}}\right) _{\ast }\otimes id\right) \pi
_{1}^{+}\left( h_{1}^{+}\right) & \left( \mathrm{ad}_{Z_{1}^{+}}^{1}\right)
^{\otimes 2}\left( L_{h_{+}^{-1}}\right) _{\ast }^{\otimes 2}\pi
_{1}^{+}\left( h_{1}^{+}\right)%
\end{array}%
\right)  \label{PL bivector}
\end{equation}%
Let us introduce the map $\pi _{+}^{2R}:H_{+}^{2}\longrightarrow \mathfrak{h}%
_{+}^{2}\otimes \mathfrak{h}_{+}^{2}$ by composing the PL bivector, regarded
as a section of $T^{\otimes 2}H_{+}^{2}$, with the right translation to the
neutral element 
\begin{equation}
\pi _{+}^{2R}\left( h_{1}^{+},Z_{1}^{+}\right) =\left( R_{\left(
h_{1}^{+},Z_{1}^{+}\right) ^{-1}}\right) _{\ast }^{\otimes 2}\pi
_{+}^{2}\left( h_{1}^{+},Z_{1}^{+}\right)  \label{PI R}
\end{equation}%
The differential of this map at the neutral element, namely $\delta :=\left(
\pi _{+}^{2R}\right) _{\ast e}:\mathfrak{h}_{+}^{2}\longrightarrow \mathfrak{%
h}_{+}^{2}\otimes \mathfrak{h}_{+}^{2}$, is a linear map$\mathfrak{\ }$that
supplies $\left( \mathfrak{h}_{+}^{2}\right) ^{\ast }$ with the Lie algebra
structure%
\begin{equation*}
\left\langle \left[ \eta _{2}^{+},\xi _{2}^{+}\right] ,X_{2}^{+}\right%
\rangle =\left\langle \eta _{2}^{+}\otimes \xi _{2}^{+},\delta \left(
X_{2}^{+}\right) \right\rangle
\end{equation*}%
for $\eta _{2}^{+},\xi _{2}^{+}\in \left( \mathfrak{h}_{+}^{2}\right) ^{\ast
}$ and $X_{2}^{+}\in \mathfrak{h}_{+}^{2}$. The Jacobi property of the Lie
bracket in $\left( \mathfrak{h}_{+}^{2}\right) ^{\ast }$ is warranted by
requiring that $\delta $ be a cocyle. Then using the expression $\left( \ref%
{PL bivector}\right) $ we get%
\begin{equation*}
\pi _{+}^{2R}\left( h_{1}^{+},Z_{1}^{+}\right) =\left( 
\begin{array}{cc}
0 & I \\ 
I & \left( \mathrm{ad}_{\mathrm{Ad}_{h_{1}^{+}}^{1}Z_{1}^{+}}^{1}\right)
^{\otimes 2}%
\end{array}%
\right) \pi _{+}^{1R}\left( h_{1}^{+}\right)
\end{equation*}%
An important case happens whenever $\delta $ is a coboundary, giving rise to
the $r$-matrix approach to integrable system \cite{RSTS 0}. In this
framework, we may see that in general a coboundary at the level $1$ fails in
to produce a coboundary at level $2$.

\subsection{Dressing vectors}

We can think of the Poisson-Lie bivector on $H_{2}^{+}$ as providing a
linear map from $T^{\ast }H_{2}^{+}$ to $TH_{2}^{+}$. It is well known that
this map defines a Lie algebra antihomomorphism from $\left( \mathfrak{h}%
_{2}^{+}\right) ^{\ast }$ to $\mathfrak{X}\left( H_{2}^{+}\right) $, the Lie
algebra of vector fields on $H_{2}^{+}$. In can be translated to a Lie
algebra antihomomorphism from $\mathfrak{h}_{2}^{-}$ to $\mathfrak{X}\left(
H_{2}^{+}\right) $ by regarding the bijection $\gamma :\mathfrak{h}%
_{2}^{-}\longrightarrow \left( \mathfrak{h}_{2}^{+}\right) ^{\ast }$ in such
a way that, for $h_{1}^{+}\in H_{2}^{+}$ and $X_{1}^{-}\in \mathfrak{h}%
_{2}^{-}$, the linear map%
\begin{equation*}
\left( X_{1}^{-}\right) ^{R}:=\left( R_{h_{+}^{-1}}\right) _{\ast
}X_{1}^{-}\longmapsto \left( id\otimes \left( R_{h_{+}^{-1}}\right) ^{\ast
}\gamma \left( X_{1}^{-}\right) \right) \pi _{2}^{+}\left( h_{1}^{+}\right)
\end{equation*}%
is a Lie algebra antihomomorphism. These vector fields are the \emph{%
dressing infinitesimal generators }associated with the factorization $%
H_{2}=H_{2}^{+}H_{2}^{-}$, and the integral submanifold of this distribution
coincides with the symplectic leaves of the Poisson-Lie structure $\left( %
\ref{PL bivector}\right) $.

The \emph{dressing vector }right translated to the tangent space at the
neutral element is 
\begin{equation}
\begin{array}{l}
\left( R_{\left( h_{1}^{+},Z_{1}^{+}\right) ^{-1}}\right) _{\ast }\left(
Y_{1}^{-},W_{1}^{-}\right) _{H_{2}^{+}}\left( h_{1}^{+},Z_{1}^{+}\right) \\ 
\qquad =\left( id\otimes \gamma \left( Y_{1}^{-},W_{1}^{-}\right) \right)
\left( R_{\left( h_{1}^{+},Z_{1}^{+}\right) ^{-1}}\right) _{\ast }^{\otimes
2}\pi _{2}^{+}\left( h_{1}^{+},Z_{1}^{+}\right)%
\end{array}
\notag
\end{equation}%
Let us come back to expression $\left( \ref{Poisson Lie struc 2}\right) $
and decompose it in the summands of $\mathfrak{h}_{1}^{+}\oplus \mathfrak{h}%
_{1}^{+}$. Since $H_{1}^{+}$ is also a Poisson-Lie group, we may write the
above expression in terms of the corresponding dressing vectors:%
\begin{equation}
\left( id\otimes \left( R_{h_{+}^{-1}}\right) ^{\ast }\gamma \left(
Y_{1}^{-}\right) \right) \pi _{1}^{+}\left( h_{1}^{+}\right) =\left(
h_{1}^{+}\right) ^{Y_{1}^{-}}  \label{dressing vect in H1+}
\end{equation}%
Therefore, the above expression reduces to%
\begin{equation}
\begin{array}{l}
\left( id\otimes \gamma \left( Y_{1}^{-},W_{1}^{-}\right) \right) \left(
R_{\left( h_{1}^{+},Z_{1}^{+}\right) ^{-1}}\right) _{\ast }^{\otimes 2}\pi
_{2}^{+}\left( h_{1}^{+},Z_{1}^{+}\right) \\ 
~=\left( \left( h_{1}^{+}\right) ^{Y_{1}^{-}}\left( h_{1}^{+}\right)
^{-1}\right. \\ 
~\qquad ,\left. \left( h_{1}^{+}\right) ^{\left( W_{1}^{-}-\left[ \mathrm{Ad}%
_{h_{1}^{+}}^{1}Z_{1}^{+},Y_{1}^{-}\right] \right) }\left( h_{1}^{+}\right)
^{-1}+\left[ \mathrm{Ad}_{h_{1}^{+}}^{1}Z_{1}^{+},\left( h_{1}^{+}\right)
^{Y_{1}^{-}}\left( h_{1}^{+}\right) ^{-1}\right] \right)%
\end{array}
\notag
\end{equation}%
To obtain the dressing infinitesimal generators, this element of $\mathfrak{h%
}_{2}^{+}$ has to be right translated to $\left( h_{1}^{+},Z_{1}^{+}\right)
\in H_{2}^{+}$ using that%
\begin{equation*}
(R_{\left( h_{1}^{+},Z_{1}^{+}\right) }^{\bullet })_{\ast \left( e,0\right)
}\left( v,Z\right) =\left( \left( R_{h_{1}^{+}}\right) _{\ast }v,\mathrm{Ad}%
_{h_{+}^{-1}}^{1}Z\right) _{(h_{1}^{+},Z_{1}^{+})}
\end{equation*}%
thus we have obtain the assignment $\left( Y_{1}^{-},W_{1}^{-}\right) \in 
\mathfrak{h}_{2}^{+}\longrightarrow \left( Y_{1}^{-},W_{1}^{-}\right)
_{H_{2}^{+}}\in \mathfrak{X}\left( H_{2}^{+}\right) $ defined as 
\begin{equation}
\left( Y_{1}^{-},W_{1}^{-}\right) _{H_{2}^{+}}\left(
h_{1}^{+},Z_{1}^{+}\right) =\left( \left( h_{1}^{+}\right) ^{Y_{1}^{-}},\Pi
_{+}\left( \mathrm{Ad}_{h_{+}^{-1}}^{1}W_{1}^{-}+\left[ \left(
Y_{1}^{-}\right) ^{h_{1}^{+}},Z_{1}^{+}\right] \right) \right)
\label{dressing vect in H2+ 2}
\end{equation}

This result can be derived also from the factorization of the product $%
\left( h_{1}^{-},X_{1}^{-}\right) \bullet \left( h_{1}^{+},X_{1}^{+}\right)
\in H_{2}$ as%
\begin{equation*}
\left( h_{1}^{-},X_{1}^{-}\right) \bullet \left( h_{1}^{+},X_{1}^{+}\right)
=\left( h_{1}^{+},X_{1}^{+}\right) ^{\left( h_{1}^{-},X_{1}^{-}\right)
}\bullet \left( h_{1}^{-},X_{1}^{-}\right) ^{\left(
h_{1}^{+},X_{1}^{+}\right) }
\end{equation*}%
where $\left( h_{1}^{+},X_{1}^{+}\right) ^{\left( h_{1}^{-},X_{1}^{-}\right)
}\in H_{1}^{+}\circledS \mathfrak{h}_{1}^{+}$ and $\left(
h_{1}^{-},X_{1}^{-}\right) ^{\left( h_{1}^{+},X_{1}^{+}\right) }\in
H_{1}^{-}\circledS \mathfrak{h}_{1}^{-}$. Hence, since%
\begin{eqnarray*}
\left( h_{1}^{-},X_{1}^{-}\right) \bullet \left( h_{1}^{+},X_{1}^{+}\right)
&=&\left( h_{1}^{-}h_{1}^{+},\mathrm{Ad}_{h_{+}^{-1}}^{1}X_{1}^{-}+X_{1}^{+}%
\right) \\
&=&\left( \left( h_{1}^{+}\right) ^{h_{1}^{-}}\left( h_{1}^{-}\right)
^{h_{1}^{+}},\mathrm{Ad}_{\left( h_{1}^{+}\right)
^{-1}}^{1}X_{1}^{-}+X_{1}^{+}\right)
\end{eqnarray*}%
we get%
\begin{equation*}
\left( h_{1}^{+},X_{1}^{+}\right) ^{\left( h_{1}^{-},X_{1}^{-}\right)
}=\left( \left( h_{1}^{+}\right) ^{h_{1}^{-}},\Pi _{+}\mathrm{Ad}_{\left(
h_{1}^{-}\right) ^{h_{1}^{+}}}^{1}\left( \mathrm{Ad}_{\left(
h_{1}^{+}\right) ^{-1}}^{1}X_{1}^{-}+X_{1}^{+}\right) \right)
\end{equation*}%
and%
\begin{equation*}
\left( h_{1}^{-},X_{1}^{-}\right) ^{\left( h_{1}^{+},X_{1}^{+}\right)
}=\left( \left( h_{1}^{-}\right) ^{h_{1}^{+}},\mathrm{Ad}_{\left( \left(
h_{1}^{-}\right) ^{h_{1}^{+}}\right) ^{-1}}^{1}\Pi _{-}\mathrm{Ad}_{\left(
h_{1}^{-}\right) ^{h_{1}^{+}}}^{1}\left( \mathrm{Ad}_{\left(
h_{1}^{+}\right) ^{-1}}^{1}X_{1}^{-}+X_{1}^{+}\right) \right)
\end{equation*}%
which define the reciprocal \emph{dressing actions} between $%
H_{1}^{+}\circledS \mathfrak{h}_{1}^{+}$ and $H_{1}^{-}\circledS \mathfrak{h}%
_{1}^{-}$. Observe that 
\begin{equation*}
\left( h_{1}^{+},X_{1}^{+}\right) ^{\left( h_{1}^{-},X_{1}^{-}\right)
\bullet \left( g_{1}^{-},Y_{1}^{-}\right) }=\left( \left(
h_{1}^{+},X_{1}^{+}\right) ^{\left( g_{1}^{-},Y_{1}^{-}\right) }\right)
^{\left( h_{1}^{-},X_{1}^{-}\right) }
\end{equation*}%
showing that its a \emph{left action}. Also, it is easy to verify that%
\begin{equation*}
\left( h_{1}^{-},X_{1}^{-}\right) ^{\left( h_{1}^{+},X_{1}^{+}\right)
\bullet \left( g_{1}^{+},Y_{1}^{+}\right) }=\left( \left(
h_{1}^{-},X_{1}^{-}\right) ^{\left( h_{1}^{+},X_{1}^{+}\right) }\right)
^{\left( g_{1}^{+},Y_{1}^{+}\right) }
\end{equation*}%
turning it into a \emph{right action}.

The infinitesimal action is attained by considering $\left(
X_{1}^{-},Y_{1}^{-}\right) \in \mathfrak{h}_{1}^{-}\circledS \mathfrak{h}%
_{1}^{-}$ and the exponential curve $\mathrm{Exp}^{\bullet }\left( t\left(
X_{1}^{-},Y_{1}^{-}\right) \right) $ to get%
\begin{equation}
\left( X_{1}^{-},Y_{1}^{-}\right) _{H_{2}^{+}}\left(
h_{1}^{+},Z_{1}^{+}\right) =\left( \left( h_{1}^{+}\right) ^{X_{1}^{-}},\Pi
_{+}\left( \mathrm{Ad}_{\left( h_{1}^{+}\right) ^{-1}}^{1}Y_{1}^{-}+\mathrm{%
ad}_{\left( X_{1}^{-}\right) ^{h_{1}^{+}}}^{1}Z_{1}^{+}\right) \right)
\label{inf h- on h+}
\end{equation}%
that coincides with the expression obtained in eq. $\left( \ref{dressing
vect in H2+ 2}\right) $.

Analogously, for the reciprocal action $\left( h_{1}^{-},Z_{1}^{-}\right)
^{\left( h_{1}^{+},X_{1}^{+}\right) }$, 
\begin{equation*}
\left( X_{1}^{+},Y_{1}^{+}\right) _{H_{2}^{-}}\left(
h_{1}^{-},Z_{1}^{-}\right) =\left( \left( h_{1}^{-}\right) ^{X_{1}^{+}},%
\mathrm{Ad}_{\left( h_{1}^{-}\right) ^{-1}}^{1}\Pi _{-}\mathrm{Ad}%
_{h_{1}^{-}}^{1}\left( Y_{1}^{+}-\mathrm{ad}_{X_{1}^{+}}^{1}Z_{1}^{-}\right)
\right)
\end{equation*}%
Moreover, they are Lie algebra morphisms:%
\begin{eqnarray*}
\left[ \left( X_{1}^{-},Y_{1}^{-}\right) _{H_{2}^{+}},\left( X_{-}^{\prime
},Y_{-}^{\prime }\right) _{H_{2}^{+}}\right] &=&-\left( \left[ \left(
X_{1}^{-},Y_{1}^{-}\right) ,\left( X_{-}^{\prime },Y_{-}^{\prime }\right) %
\right] \right) _{H_{2}^{+}} \\
&& \\
\left[ \left( X_{1}^{+},Y_{1}^{+}\right) _{H_{2}^{-}},\left( X_{+}^{\prime
},Y_{+}^{\prime }\right) _{H_{2}^{-}}\right] &=&\left( \left[ \left(
X_{1}^{+},Y_{1}^{+}\right) ,\left( X_{+}^{\prime },Y_{+}^{\prime }\right) %
\right] \right) _{H_{2}^{-}}
\end{eqnarray*}

\subsection{Hamiltonian dressing action}

Besides the Poisson-Lie structure on $H_{2}^{+}=H_{1}^{+}\circledS \mathfrak{%
h}_{1}^{+}$ described in $\left( \ref{PL structure}\right) $, one may take
profit from the identification $H_{1}^{+}\times \mathfrak{h}_{1}^{+}\cong
TH_{1}^{+}$ to supply $H_{2}^{+}$ with a symplectic structure borrowed from
the canonical one in $T^{\ast }H_{1}^{+}$. In fact, let us assume that $%
H_{1} $ is equipped with nondegenerate symmetric\textit{\ }bilinear form $%
\left( ,\right) _{1}:\mathfrak{h}_{1}\otimes \mathfrak{h}_{1}\longrightarrow 
\mathbb{R}$ such that its restriction to $\mathfrak{h}_{1}^{+}$ is also
nondegenerate, and let\textit{\ }$\sigma :\mathfrak{h}_{1}\longrightarrow 
\mathfrak{h}_{1}^{\ast }\ $be the induced the linear bijection. Then,\textit{%
\ }$H_{2}^{+}$ is a symplectic manifold with the symplectic form $\omega
_{1}^{+}$ defined as 
\begin{equation*}
\begin{array}{l}
\mathbf{\langle }\omega _{1}^{+},(v_{1}^{+},X_{1}^{+})\otimes
(w_{1}^{+},Y_{1}^{+})\mathbf{\rangle }_{(h_{1}^{+},Z_{1}^{+})} \\ 
\qquad \qquad =-\left( X_{1}^{+},\left( h_{1}^{+}\right)
^{-1}w_{1}^{+}\right) _{1}+\left( Y_{1}^{+},\left( h_{1}^{+}\right)
^{-1}v_{1}^{+}\right) _{1} \\ 
\qquad \qquad \qquad +\left( Z_{1}^{+},[\left( h_{1}^{+}\right)
^{-1}v_{1}^{+},\left( h_{1}^{+}\right) ^{-1}w_{1}^{+}]\right) _{1}%
\end{array}%
\end{equation*}%
for $(h_{1}^{+},Z_{1}^{+})\in H_{1}^{+}\times \mathfrak{h}_{1}^{+}$ and $%
(v_{1}^{+},X_{1}^{+}),(w_{1}^{+},Y_{1}^{+})\in T_{h_{+}}H_{1}^{+}\oplus 
\mathfrak{h}_{1}^{+}$.

We now seek for a hamiltonian function associated with the infinitesimal
generators of the dressing actions of $H_{2}^{-}$ on $H_{2}^{+}$, given in
eq. $\left( \ref{inf h- on h+}\right) $, relative to the above symplectic
structure. In doing so, for a vector $\left( X_{1}^{-},Y_{1}^{-}\right) \in 
\mathfrak{h}_{2}^{-}$, we split the action as the composition of actions of $%
\left( X_{1}^{-},0\right) $ followed by the action of $\left(
0,Y_{1}^{-}\right) $.

From the expression of the infinitesimal generator $\left( \ref{dressing
vect in H2+ 2}\right) $ and using the relations $\left( \ref{R1},\ref{R2}%
\right) $, we write down\ 
\begin{equation}
\left( X_{1}^{-},0\right) _{H_{2}^{+}}\left( h_{1}^{+},Z_{1}^{+}\right)
=\left( h_{1}^{+}\Pi _{+}\mathrm{Ad}_{\left( h_{1}^{+}\right)
^{-1}}^{1}X_{1}^{-},\Pi _{+}\left[ \Pi _{-}\mathrm{Ad}_{\left(
h_{1}^{+}\right) ^{-1}}^{1}X_{1}^{-},Z_{1}^{+}\right] \right)
\label{dressing vect in H2+ 1d}
\end{equation}%
This vector field coincides with the lift of the infinitesimal generators of
the dressing action of $H_{1}^{-}$ on $H_{1}^{+}$ to the tangent bundle $%
TH_{1}^{+}$. These vector fields are hamiltonian provided the bilinear form $%
\left( ,\right) _{1}$ is also $\mathit{\mathrm{Ad}}^{1}$-invariant and, in
such a case, the Hamilton function $\theta _{X_{1}^{-}}$ associated with $%
X_{1}^{-}\in \mathfrak{h}_{1}^{+}$ is \textit{\ }%
\begin{equation*}
\theta _{X_{1}^{-}}(h_{1}^{+},Z_{1}^{+})=\left( Z_{1}^{+},\Pi _{+}\mathrm{Ad}%
_{\left( h_{1}^{+}\right) ^{-1}}X_{1}^{-}\right) _{1}=\left\langle \Theta
(h_{1}^{+},Z_{1}^{+}),X_{1}^{-}\right\rangle
\end{equation*}%
where $\Theta :H_{+}^{2}\longrightarrow \left( \mathfrak{h}_{1}^{-}\right)
^{\ast }$ is the $\mathrm{Ad}^{1}$-equivariant momentum map 
\begin{equation}
\Theta (h_{1}^{+},Z_{1}^{+})=\gamma \left( \Pi _{+}\mathrm{Ad}_{\left(
h_{1}^{+}\right) ^{-1}}\gamma ^{-1}\left( \sigma \left( Z_{1}^{+}\right)
\right) \right)  \label{momentum map 0}
\end{equation}

The remaining term in the infinitesimal generator $\left( \ref{dressing vect
in H2+ 2}\right) $ is 
\begin{equation*}
\left( 0,W_{1}^{-}\right) _{H_{+}^{2}}\left( h_{+},Z_{+}\right) =\left(
0,\Pi _{+}\mathrm{Ad}_{\left( h_{1}^{+}\right) ^{-1}}W_{1}^{-}\right)
\end{equation*}%
Observe that%
\begin{equation*}
\imath _{\left( 0,W_{1}^{-}\right) _{H_{+}^{2}}}\omega _{1}^{+}=-\left(
h_{1}^{+}\right) ^{-1}\sigma \left( \left( h_{1}^{+}\right) ^{-1}\left(
h_{1}^{+}\right) ^{W_{1}^{-}}\right)
\end{equation*}%
and the condition $d\imath _{\left( 0,W_{1}^{-}\right) _{H_{+}^{2}}}\omega
_{1}^{+}=0$ imposes strong restrictions on $\sigma $ so, in general, the Lie
derivative $\mathbf{L}_{\left( 0,W_{1}^{-}\right) _{H_{+}^{2}}}\omega
_{1}^{+}\neq 0$ meaning that the infinitesimal transformation induced by the
vector field $\left( 0,W_{1}^{-}\right) _{H_{+}^{2}}$ is not a
symplectomorphism. Related to this fact, one may see that the left action of 
$H_{2}$ on itself is in general not hamiltonian relative to the symplectic
form of $H_{1}\times \mathfrak{h}_{1}$.

In the following sections we study some integrable systems on these phase
spaces with flows described by dressing vector field like $\left( \ref%
{dressing vect in H2+ 1d}\right) $.

\section{Phase spaces in $H_{2}$}

We now consider $H_{2}$ as phase space equipped with the nondegenerate
Poisson bracket inherited from the canonical symplectic form of $T^{\ast
}H_{1}$ and, following the Dirac procedure developed in ref. \cite{CapMon
JMP}, we construct a class of phase subspaces of $H_{2}$ on which nontrivial
integrable systems naturally arise.

The nondegenerate Poisson bracket we consider here is obtained in analogous
way as in the previous subsection, by using the inner product $\left(
,\right) _{1}$ and the induced bijection $\sigma :\mathfrak{h}%
_{1}\longrightarrow \mathfrak{h}_{1}^{\ast }$, with its restrictions $\left.
\sigma \right\vert \mathfrak{h}_{1}^{\pm }:\mathfrak{h}_{1}^{\pm
}\longrightarrow \mathfrak{h}_{1}^{\pm \ast }$ being also linear bijections.
Thus the symplectic structure in $H_{2}$ is 
\begin{equation}
\mathbf{\langle }\omega _{2},(v,X)\otimes (w,Y)\mathbf{\rangle }%
_{(h_{1},Z_{1})}=-\left( X,h_{1}^{-1}w\right) _{1}+\left(
Y,h_{1}^{-1}v\right) _{1}+\left( Z_{1},[h_{1}^{-1}v,h_{1}^{-1}w]\right) _{1}
\label{sympl form on TH1}
\end{equation}%
for $(h_{1},Z_{1})\in H_{1}\times \mathfrak{h}_{1}$, $(v,X),(w,Y)\in
T_{(h_{1},Z_{1})}\left( H_{1}\times \mathfrak{h}_{1}\right) $. Let $\mathcal{%
F},\mathcal{H}$ be functions on $H_{1}\times \mathfrak{h}_{1}$ and let us
write their differential as $d\mathcal{F}=\left( \mathbf{d}\mathcal{F}%
,\delta \mathcal{F}\right) \in T^{\ast }H_{1}\oplus \mathfrak{h}_{1}^{\ast }$%
, then the associated Poisson bracket is 
\begin{eqnarray}
\left\{ \mathcal{F},\mathcal{H}\right\} _{2}\left( h_{1},Z_{1}\right)
&=&\left\langle \mathbf{d}\mathcal{F},h\sigma ^{-1}\left( \delta \mathcal{H}%
\right) \right\rangle -\left\langle \mathbf{d}\mathcal{H},h\sigma
^{-1}\left( \delta \mathcal{F}\right) \right\rangle  \label{PB in Hxh} \\
&&-\left\langle \sigma \left( Z_{1}\right) ,[\sigma ^{-1}\left( \delta 
\mathcal{F}\right) ,\sigma ^{-1}\left( \delta \mathcal{H}\right)
]\right\rangle  \notag
\end{eqnarray}%
From this expression we get the hamiltonian vector field of $\mathcal{H}$ 
\begin{equation*}
V_{\mathcal{H}}\left( h_{1},Z_{1}\right) =\left( h_{1}\sigma ^{-1}\left(
\delta \mathcal{H}\right) ,\sigma ^{-1}\left( \mathrm{ad}_{\sigma
^{-1}\left( \delta \mathcal{H}\right) }^{1\ast }\sigma \left( Z_{1}\right) -h%
\mathbf{d}\mathcal{H}\right) \right)
\end{equation*}%
and the Poisson bracket $\left\{ \mathcal{F},\mathcal{H}\right\} _{2}$ just
means the Lie derivative of the function $\mathcal{F}$ along the the vector
field $V_{\mathcal{H}}$.

The main idea of the approach to integrability we use here is to obtain non
trivial integrable systems as the reduction of an almost trivial system
defined in a phase space to some submanifold. It fits perfectly in the realm
of Dirac method for constrained systems since it produces Lie derivatives of
functions on the whole phase space along the projection of the hamiltonian
vector fields on the tangent space of the constrained submanifold, giving
rise to a representation of these vectors fields in terms of the geometrical
data in the total phase space. Integral curves of this projected vector
field are the trajectories of the constrained hamiltonian system. In the
rest of the current section we adapt the approach developed in reference $%
\cite{CapMon JMP}$ to the framework of semidirect product and factorization
as introduced above.

\subsection{Fibration of symplectic submanifolds in $H_{2}$}

The starting point is the phase space $\left( H_{2},\left\{ ,\right\}
_{2}\right) $, and we introduce the fibration 
\begin{eqnarray}
\Psi _{2}:H_{1}\times \mathfrak{h}_{1} &\longrightarrow &H_{1}^{-}\times 
\mathfrak{h}_{1}^{-}  \label{proyeccion-1} \\
\left( h_{1},X_{1}\right) &\longmapsto &\left( h_{1}^{-},X_{1}^{-}\right) 
\notag
\end{eqnarray}%
such that the fiber on $\left( h_{1}^{-},X_{1}^{-}\right) \in
H_{1}^{-}\times \mathfrak{h}_{1}^{-}$ is described as 
\begin{equation*}
\mathcal{N}_{2}\left( h_{1}^{-},X_{1}^{-}\right) =\Psi _{2}^{-1}\left(
h_{1}^{-},X_{1}^{-}\right) =\left\{ \left(
h_{1}^{+}h_{1}^{-},X_{1}^{+}+X_{1}^{-}\right) /h_{1}^{+}\in
H_{1}^{+},~X_{1}^{+}\in \mathfrak{h}_{1}^{+}\right\}
\end{equation*}%
In particular, $\mathcal{N}_{2}\left( e,0\right) =H_{1}^{+}\times \mathfrak{h%
}_{1}^{+}$. Each fiber $\mathcal{N}_{2}\left( h_{1}^{-},X_{1}^{-}\right) $
can be supplied with a nondegenerate Dirac bracket constructed following
ref. \cite{CapMon JMP}. We shall use the linear bijection $\sigma :\mathfrak{%
h}_{1}\longrightarrow \mathfrak{h}_{1}^{\ast }$ allows to translate those
results to the current framework.

In order to simplify the notation we introduce the projectors $\mathbb{A}%
_{\pm }^{k}\left( h\right) $, $k=1,2$, defined as%
\begin{equation*}
\mathbb{A}_{\pm }^{k}\left( h\right) :=\mathrm{Ad}_{h^{-1}}^{k}\Pi _{\pm }%
\mathrm{Ad}_{h}^{k}
\end{equation*}%
such that 
\begin{equation*}
\left\{ 
\begin{array}{l}
\mathbb{A}_{\pm }^{k}\left( h\right) \mathbb{A}_{\pm }^{k}\left( h\right) =%
\mathbb{A}_{\pm }^{k}\left( h\right) \\ 
\mathbb{A}_{\mp }^{k}\left( h\right) \mathbb{A}_{\pm }^{k}\left( h\right) =0
\\ 
\mathbb{A}_{+}^{k}\left( h\right) +\mathbb{A}_{-}^{k}\left( h\right) =Id%
\end{array}%
\right.
\end{equation*}%
which will be used in the following.

Therefore the Dirac bracket on $\mathcal{N}_{2}\left(
h_{1}^{-},X_{1}^{-}\right) $ for $\mathcal{F},\mathcal{H}\in C^{\infty
}\left( H_{1}\times \mathfrak{h}_{1}\right) $ is 
\begin{eqnarray}
\left\{ \mathcal{F},\mathcal{H}\right\} _{2}^{D}\left( h_{1},Z_{1}\right)
&=&\left\langle h_{1}\mathbf{d}\mathcal{F},\mathbb{A}_{+}^{1}\left(
h_{1}^{-}\right) \sigma ^{-1}\left( \delta \mathcal{H}\right) \right\rangle
\label{Dirac bracket G+xg+  I} \\
&&-\left\langle h_{1}\mathbf{d}\mathcal{H},\mathbb{A}_{+}^{1}\left(
h_{1}^{-}\right) \sigma _{1}^{-1}\left( \delta \mathcal{F}\right)
\right\rangle  \notag \\
&&-\left\langle \sigma \left( Z_{1}\right) ,[\mathbb{A}_{+}^{1}\left(
h_{1}^{-}\right) \sigma ^{-1}\left( \delta \mathcal{F}\right) ,\mathbb{A}%
_{+}^{1}\left( h_{1}^{-}\right) \sigma ^{-1}\left( \delta \mathcal{H}\right)
]\right\rangle  \notag
\end{eqnarray}%
where we have denoted the differential of a function $\mathcal{F}$ as $d%
\mathcal{F}=\left( \mathbf{d}\mathcal{F},\delta \mathcal{F}\right) $ for its
components in $T_{\left( h_{1},Z_{1}\right) }^{\ast }\left( H_{1}\times 
\mathfrak{h}_{1}\right) =T_{h_{1}}^{\ast }H_{1}\times \mathfrak{h}_{1}^{\ast
}$. In the particular case $h_{1}^{-}=e$, $\eta _{1}^{-}=0$, we have $%
\mathcal{N}_{2}\left( e,0\right) =H_{2}^{+}$ and the Dirac bracket reduces to%
\begin{eqnarray*}
\left\{ \mathcal{F},\mathcal{H}\right\} _{2}^{D}\left(
h_{1}^{+},Z_{1}^{+}\right) &=&\left\langle h_{1}^{+}\mathbf{d}\mathcal{F}%
,\Pi _{+}\sigma ^{-1}\left( \delta \mathcal{H}\right) \right\rangle
-\left\langle h_{1}^{+}\mathbf{d}\mathcal{H},\Pi _{+}\sigma ^{-1}\left(
\delta \mathcal{F}\right) \right\rangle \\
&&-\left( Z_{1}^{+},[\Pi _{+}\sigma ^{-1}\left( \delta \mathcal{F}\right)
,\Pi _{+}\sigma ^{-1}\left( \delta \mathcal{H}\right) ]\right) _{1}
\end{eqnarray*}%
as expected.

Let us consider a generic hamiltonian function $\mathcal{H}$ on $H_{1}\times 
\mathfrak{h}_{1}$ so, from the Poisson-Dirac bracket $\left( \ref{Dirac
bracket G+xg+ I}\right) $ one may obtain the corresponding hamiltonian
vector field which turns to be 
\begin{eqnarray}
&&V_{\mathcal{H}}^{\mathcal{N}}\left( h,Z\right)  \label{ham vect field} \\
&=&\left( h\left( \mathbb{A}_{+}^{1}\left( h_{1}^{-}\right) \sigma
^{-1}\left( \delta \mathcal{H}\right) \right) \right. ,  \notag \\
&&,\left. \sigma ^{-1}\left( \gamma \left( \mathbb{A}_{-}^{1}\left(
h_{1}^{-}\right) \left( [\gamma ^{-1}\left( \sigma \left( Z\right) \right) ,%
\mathbb{A}_{+}^{1}\left( h_{1}^{-}\right) \sigma ^{-1}\left( \delta \mathcal{%
H}\right) ]-\gamma ^{-1}\left( h\mathbf{d}\mathcal{H}\right) \right) \right)
\right) \right)  \notag
\end{eqnarray}%
The Hamilton equation are then 
\begin{equation}
\left\{ 
\begin{array}{l}
h_{1}^{-1}\dot{h}_{1}=\mathbb{A}_{+}^{1}\left( h_{1}^{-}\right) \sigma
^{-1}\left( \delta \mathcal{H}\right) \\ 
\\ 
\dot{Z}_{1}=\sigma ^{-1}\left( \gamma \left( \mathbb{A}_{-}^{1}\left(
h_{1}^{-}\right) \left( [\gamma ^{-1}\left( \sigma \left( Z\right) \right) ,%
\mathbb{A}_{+}^{1}\left( h_{1}^{-}\right) \sigma ^{-1}\left( \delta \mathcal{%
H}\right) ]-\gamma ^{-1}\left( h\mathbf{d}\mathcal{H}\right) \right) \right)
\right)%
\end{array}%
\right.  \label{Ham eqs on N(g-,eta-)}
\end{equation}

It is interesting to specialize this structures to left invariant functions.
Left translations on $H_{1}$ lifted to $TH_{1}=H_{1}\times \mathfrak{h}_{1}$
are 
\begin{equation*}
\left( L_{g_{1}}\right) _{\ast }(h_{1},Z_{1})=(g_{1}h_{1},Z_{1})
\end{equation*}%
meaning that left invariant functions are those satisfying $\mathcal{F}%
_{L}\left( g_{1}h_{1},Z_{1}\right) =\mathcal{F}_{L}\left( h_{1},Z_{1}\right) 
$, and the Dirac bracket of two left invariant functions $\mathcal{F}_{L},%
\mathcal{H}_{L}$ reduces to 
\begin{equation*}
\left\{ \mathcal{F}_{L},\mathcal{H}_{L}\right\} _{2}^{D}\left(
h_{1},Z_{1}\right) =-\left( Z_{1},[\mathbb{A}_{+}^{1}\left( h_{1}^{-}\right)
\sigma ^{-1}\left( \delta \mathcal{F}_{L}\right) ,\mathbb{A}_{+}^{1}\left(
h_{1}^{-}\right) \sigma ^{-1}\left( \delta \mathcal{H}_{L}\right) ]\right)
_{1}
\end{equation*}%
At $h_{1}^{-}=e$ and $\eta _{1}^{-}=0$, it reduces to%
\begin{equation}
\left\{ \mathcal{F}_{L},\mathcal{H}_{L}\right\} _{2}^{D}\left(
h_{1}^{+},Z_{1}^{+}\right) =-\left( Z_{1}^{+},[\Pi _{+}\sigma ^{-1}\left(
\delta \mathcal{F}_{L}\right) ,\Pi _{+}\sigma ^{-1}\left( \delta \mathcal{H}%
_{L}\right) ]\right) _{1}  \label{DP - LP bracket}
\end{equation}%
that is the standard \emph{Lie-Poisson bracket} on $\mathfrak{h}_{1}^{+}$.
The hamiltonian vector field for left invariant functions is 
\begin{equation*}
V_{\mathcal{H}_{L}}\left( h,Z\right) =\gamma \left( \mathbb{A}_{-}^{1}\left(
h_{1}^{-}\right) \left[ \mathbb{A}_{+}^{1}\left( h_{1}^{-}\right) \sigma
^{-1}\left( \delta \mathcal{H}_{L}\right) ,\gamma ^{-1}\left( \sigma \left(
Z\right) \right) \right] \right)
\end{equation*}

\subsection{The left action of $H_{1}$ on $H_{2}$ and its restriction to $%
\mathcal{N}\left( h_{1}^{-},\protect\eta _{1}^{-}\right) $}

In \emph{body coordinates} for $TH_{1}$, the \emph{left translation }is\emph{%
\ }$\left( L_{g_{1}}\right) _{\ast }(h_{1},Z_{1})=(g_{1}h_{1},Z_{1})$, with
the infinitesimal generator 
\begin{equation*}
X_{H_{2}}(h_{1},Z_{1}):=\left. \frac{d}{dt}(e^{tX_{1}}h_{1},Z_{1})\right%
\vert _{t=0}=(X_{1}h_{1},0)
\end{equation*}%
This action is hamiltonian referred to the symplectic form $\left( \ref%
{sympl form on TH1}\right) $ with associated Ad-equivariant momentum map $%
\Phi _{B}:H_{2}\longrightarrow \mathfrak{h}_{1}^{\ast }$ 
\begin{equation}
\Phi _{B}(h_{1},Z_{1})=\gamma \left( \mathrm{Ad}_{h_{1}}^{1}\gamma
^{-1}\left( \sigma \left( Z_{1}\right) \right) \right)
\label{left translation Hxh mom map}
\end{equation}%
Hence, the momentum functions 
\begin{equation}
\phi _{X_{1}}(h_{1},Z_{1}):=\left\langle \gamma \left( \mathrm{Ad}%
_{h_{1}}^{1}\gamma ^{-1}\left( \sigma \left( Z_{1}\right) \right) \right)
,X_{1}\right\rangle =\left( Z_{1},\mathrm{Ad}_{h_{1}^{-1}}^{1}X_{1}\right)
_{1}  \label{left translation Hxh mom fun}
\end{equation}%
produce the Lie derivative of a function $\mathcal{F}$ on $H_{2}$ along the
projection of the vector field $X_{H_{2}}$ on $\mathcal{N}_{2}\left(
h_{1}^{-},\eta _{1}^{-}\right) $ through the Dirac bracket $\left( \ref%
{Dirac bracket G+xg+ I}\right) $. The differential of the momentum function
is%
\begin{equation}
\left( \mathbf{d}\phi _{X},\delta \phi _{X}\right) =\left( h_{1}^{-1}\gamma
\left( \left[ \gamma ^{-1}\left( \sigma \left( Z_{1}\right) \right) ,\mathrm{%
Ad}_{h_{1}^{-1}}^{1}X_{1}\right] \right) ,\sigma \left( \mathrm{Ad}%
_{h_{1}^{-1}}^{1}X_{1}\right) \right)  \label{dif moment funct}
\end{equation}%
and the Lie derivative of a function $\mathcal{F}$ along the infinitesimal
generator $X_{H_{2}}$ is given by the Dirac bracket 
\begin{eqnarray*}
&&\left\{ \mathcal{F},\phi _{X}\right\} _{2}^{D}\left( h_{1},Z_{1}\right) \\
&=&\left\langle \mathbf{d}\mathcal{F},h_{1}\mathbb{A}_{+}^{1}\left(
h_{1}^{-}\right) X_{1}\right\rangle \\
&&-\left\langle \delta \mathcal{F},\sigma ^{-1}\left( \gamma \left( \mathbb{A%
}_{-}^{1}\left( h_{1}^{-}\right) \left[ \gamma ^{-1}\left( \sigma \left(
Z_{1}\right) \right) ,\mathrm{Ad}_{\left( h_{1}^{-}\right) ^{-1}}\Pi _{-}%
\mathrm{Ad}_{\left( h_{1}^{+}\right) ^{-1}}X_{1}\right] \right) \right)
\right\rangle
\end{eqnarray*}

In particular, the\ Dirac bracket between momentum function $\phi _{X},\phi
_{Y}$ is%
\begin{eqnarray*}
&&\left\{ \phi _{X},\phi _{Y}\right\} _{2}^{D}\left( h_{1},Z_{1}\right) \\
&=&\phi _{\left[ X,Y\right] }\left( h_{1},Z_{1}\right) \\
&&-\left( Z_{1},\left[ \mathrm{Ad}_{\left( h_{1}^{-}\right) ^{-1}}^{1}\Pi
_{-}\mathrm{Ad}_{\left( h_{1}^{+}\right) ^{-1}}^{1}X,\mathrm{Ad}_{\left(
h_{1}^{-}\right) ^{-1}}^{1}\Pi _{-}\mathrm{Ad}_{\left( h_{1}^{+}\right)
^{-1}}^{1}Y\right] \right) _{1}
\end{eqnarray*}%
for $X,Y\in \mathfrak{h}_{1}$. Observe that the second term vanish whenever $%
\Pi _{-}^{\ast }\sigma \left( Z_{1}\right) $ \emph{is a character of} $%
\mathfrak{h}_{1}^{-}$ meaning that the set of hamiltonian vector fields $%
\left\{ V_{\phi _{X}}\right\} _{X\in \mathfrak{h}}$ forms a Lie subalgebra
of $\mathfrak{X}\left( \mathcal{N}_{2}\left( h_{1}^{-},\eta _{1}^{-}\right)
\right) $%
\begin{equation*}
\left[ V_{\phi _{X}},V_{\phi _{Y}}\right] =-V_{\phi _{\left[ X,Y\right] }}
\end{equation*}%
and defines a foliation in $\mathcal{N}_{2}\left( h_{1}^{-},Z_{1}^{-}\right) 
$. It is just in this case when the map $X\longrightarrow V_{\phi _{X}}$
defines a infinitesimal left action of $\mathfrak{h}_{1}$ on $\mathcal{N}%
_{2}\left( h_{1}^{-},Z_{1}^{-}\right) $, and they are the infinitesimal
generators of the action $H_{1}\times \mathcal{N}_{2}\left( h_{1}^{-},\eta
_{1}^{-}\right) \longrightarrow \mathcal{N}_{2}\left( h_{1}^{-},\eta
_{1}^{-}\right) $%
\begin{eqnarray*}
&&\mathsf{d}\left( g,\left( h_{1}^{+}h_{1}^{-},\eta _{1}^{+}+\eta
_{1}^{-}\right) \right) \\
&=&\left( g\mathrm{Ad}_{\left( h_{1}^{-}\right) ^{-1}}^{1}\Pi _{+}\left(
h_{1}^{+}\right) ^{-1}gh_{1}^{+}\right. , \\
&&\qquad ,\sigma ^{-1}\left( \gamma \left( \mathrm{Ad}_{\left(
h_{1}^{-}\right) ^{-1}}^{1}\Pi _{-}\mathrm{Ad}_{\Pi _{-}\left(
h_{1}^{+}\right) ^{-1}gh_{1}^{+}}^{1}\mathrm{Ad}_{h_{1}^{-}}^{1}\gamma
^{-1}\left( \sigma \left( Z_{1}\right) \right) \right) \right)
\end{eqnarray*}%
for $g\in H_{1}$.

\section{Dynamics and integrable systems on $H_{2}$}

\subsection{Hamilton equations and collective dynamics on $\mathcal{N}\left(
g_{1}^{-},\protect\eta _{1}^{-}\right) $}

The Hamilton equations of motion on $\mathcal{N}\left( g_{1}^{-},\eta
_{1}^{-}\right) $ were derived from the Dirac bracket $\left( \ref{Dirac
bracket G+xg+ I}\right) $ in eq. $\left( \ref{Ham eqs on N(g-,eta-)}\right) $%
, for any hamiltonian function $\mathcal{H}$ on $H_{1}\times \mathfrak{h}%
_{1}=H_{2}$. In order to study integrable systems, we shall consider a
general \emph{collective hamiltonian} 
\begin{equation*}
\mathcal{H}(h_{1},Z_{1}):=\mathsf{h}\left( \Phi _{B}(h_{1},Z_{1})\right)
\end{equation*}%
where $\mathsf{h}:\mathfrak{h}_{1}^{\ast }\longrightarrow \mathbb{R}$ is $%
\mathrm{Ad}$-invariant. Observe that collective hamiltonians with
Ad-invariant $\mathsf{h}:\mathfrak{h}_{1}^{\ast }\longrightarrow \mathbb{R}$
are naturally left invariant, i.e., for any $g\in H_{1}$: 
\begin{equation*}
\mathcal{H}\left( \left( L_{g}\right) _{\ast }(h,Z)\right) =\mathsf{h}\left(
\Phi _{B}\left( \left( L_{g}\right) _{\ast }(h,Z)\right) \right) =\mathsf{h}%
\left( \mathrm{Ad}_{g^{-1}}^{\ast }\Phi _{B}\left( h,Z)\right) \right) =%
\mathcal{H}\left( h,Z)\right)
\end{equation*}

Let us introduce the Legendre transform of $\mathsf{h}:\mathfrak{h}%
_{1}^{\ast }\longrightarrow \mathbb{R}$, defined as the map $\mathcal{L}_{%
\mathsf{h}}:\mathfrak{h}_{1}^{\ast }\longrightarrow \mathfrak{h}_{1}$ with 
\begin{equation*}
\mathsf{k}_{1}\left( \mathcal{L}_{\mathsf{h}}(\eta _{1}),X_{1}\right)
=\left. \frac{d}{dt}\mathsf{h}\left( \eta _{1}+t\gamma \left( X_{1}\right)
\right) \right\vert _{t=0}
\end{equation*}%
Because of the Ad-invariance of $\mathsf{h}$ we have that%
\begin{equation*}
\mathcal{L}_{\mathsf{h}}(\mathrm{Ad}_{h}^{\ast }\eta _{1})=\mathrm{Ad}%
_{h^{-1}}\mathcal{L}_{\mathsf{h}}(\eta _{1})
\end{equation*}%
so 
\begin{equation}
\left. d\mathcal{H}\right\vert _{\left( h_{1},Z_{1}\right) }=\left( 0,\sigma
\left( \mathcal{L}_{\mathsf{h}}\left( \sigma \left( Z_{1}\right) \right)
\right) \right)  \label{diff goPhi}
\end{equation}%
and the Hamilton equations $\left( \ref{Ham eqs on N(g-,eta-)}\right) $ turn
into 
\begin{equation*}
\left\{ 
\begin{array}{l}
h_{1}^{-1}\dot{h}_{1}=\mathbb{A}_{+}^{1}\left( h_{1}^{-}\right) \mathcal{L}_{%
\mathsf{h}}\left( \sigma \left( Z_{1}\right) \right) \\ 
\\ 
\dot{Z}_{1}=\sigma ^{-1}\left( \gamma \left( \mathbb{A}_{-}^{1}\left(
h_{1}^{-}\right) \left( [\gamma ^{-1}\left( \sigma \left( Z_{1}\right)
\right) ,\mathbb{A}_{+}^{1}\left( h_{1}^{-}\right) \mathcal{L}_{\mathsf{h}%
}\left( \sigma \left( Z_{1}\right) \right) ]\right) \right) \right)%
\end{array}%
\right.
\end{equation*}

It can be decomposed into the dynamical content on each factor of $%
H_{1}=H_{1}^{+}H_{1}^{-}$ and $\mathfrak{h}_{1}=\mathfrak{h}_{1}^{+}\oplus 
\mathfrak{h}_{1}^{-}$ by writing 
\begin{equation*}
h_{1}^{-1}\dot{h}_{1}=\left( h_{1}^{+}h_{1}^{-}\right) ^{-1}\frac{d}{dt}%
\left( h_{1}^{+}h_{1}^{-}\right) =\mathrm{Ad}_{\left( h_{1}^{-}\right)
^{-1}}^{1}\left( \left( h_{1}^{+}\right) ^{-1}\dot{h}_{1}^{+}+\dot{h}%
_{1}^{-}\left( h_{1}^{-}\right) ^{-1}\right)
\end{equation*}%
therefore 
\begin{equation}
\begin{array}{l}
\left\{ 
\begin{array}{l}
\left( h_{1}^{+}\right) ^{-1}\dot{h}_{1}^{+}=\Pi _{+}\mathrm{Ad}%
_{h_{1}^{-}}^{1}\mathcal{L}_{\mathsf{h}}\left( \sigma \left( Z_{1}\right)
\right) \\ 
\\ 
\dot{Z}_{1}^{+}=\sigma ^{-1}\left( \gamma \left( \mathbb{A}_{-}^{1}\left(
h_{1}^{-}\right) [\gamma ^{-1}\left( \sigma \left( Z_{1}\right) \right) ,%
\mathbb{A}_{+}^{1}\left( h_{1}^{-}\right) \mathcal{L}_{\mathsf{h}}\left(
\sigma \left( Z_{1}\right) \right) ]\right) \right)%
\end{array}%
\right. \\ 
\\ 
\left\{ 
\begin{array}{l}
\left( h_{1}^{-}\right) ^{-1}\dot{h}_{1}^{-}=0 \\ 
\\ 
\dot{Z}_{1}^{-}=0%
\end{array}%
\right.%
\end{array}
\label{collective ham eq. on N}
\end{equation}

\subsection{$\left( \mathbf{\Omega ,\Gamma }\right) $ coordinates \label%
{Gamma Omega}}

The equations $\left( \ref{collective ham eq. on N}\right) $ are easily
handled after introducing the variables%
\begin{equation}
\left\{ 
\begin{array}{c}
\mathbf{\Omega }_{1}=\mathrm{Ad}_{h_{1}^{-}}^{1}\mathcal{L}_{\mathsf{h}%
}\left( \sigma \left( Z_{1}\right) \right) \\ 
\\ 
\mathbf{\Gamma }_{1}=\mathrm{Ad}_{h_{1}^{-}}^{1}\gamma ^{-1}\left( \sigma
\left( Z_{1}\right) \right)%
\end{array}%
\right.  \label{Omega-Gamma}
\end{equation}%
which are related as%
\begin{equation}
\mathbf{\Omega }_{1}=\mathcal{L}_{\mathsf{h}}\left( \gamma \left( \mathbf{%
\Gamma }_{1}\right) \right)  \label{Omega 1}
\end{equation}%
Also, observe that%
\begin{equation*}
\Pi _{+}\mathbf{\Gamma }_{1}\mathbf{=}\Pi _{+}\mathrm{Ad}_{h_{1}^{-}}^{1}\Pi
_{+}\gamma ^{-1}\left( \sigma \left( Z_{1}\right) \right) \mathbf{=}\gamma
^{-1}\left( Ad_{\left( h_{1}^{-}\right) ^{-1}}^{\ast }\Pi _{-}^{\ast }\sigma
\left( Z_{1}\right) \right)
\end{equation*}%
and because $\Pi _{-}^{\ast }\sigma \left( Z_{1}\right) $ is a character of $%
\mathfrak{h}_{1}^{-}$, it turns in%
\begin{equation*}
\Pi _{+}\mathbf{\Gamma }_{1}\mathbf{=}\gamma ^{-1}\left( \Pi _{-}^{\ast
}\sigma \left( Z_{1}\right) \right)
\end{equation*}%
meaning that $\gamma \left( \Pi _{+}\mathbf{\Gamma }_{1}\right) =\Pi
_{-}^{\ast }\gamma \left( \mathbf{\Gamma }_{1}\right) $ is a character of $%
\mathfrak{h}_{1}^{-}$.

The equations $\left( \ref{collective ham eq. on N}\right) $ state that $%
\dot{Z}_{1}^{-}=0$, so we can write%
\begin{equation*}
\mathrm{Ad}_{h_{1}^{-}}\gamma ^{-1}\left( \sigma \left( \dot{Z}%
_{1}^{+}\right) \right) =\mathrm{Ad}_{h_{1}^{-}}\gamma ^{-1}\left( \sigma
\left( \dot{Z}_{1}\right) \right) =\mathbf{\dot{\Gamma}}_{1}
\end{equation*}%
and 
\begin{equation*}
\mathbf{\dot{\Gamma}}_{1}^{+}=\Pi _{+}\mathrm{Ad}_{h_{1}^{-}}\gamma
^{-1}\left( \sigma \left( \dot{Z}_{1}\right) \right) =\Pi _{+}\mathrm{Ad}%
_{h_{1}^{-}}\gamma ^{-1}\left( \sigma \left( \dot{Z}_{1}^{-}\right) \right)
=0
\end{equation*}%
Therefore, the collective Hamilton equations $\left( \ref{collective ham eq.
on N}\right) $ can be written as 
\begin{equation}
\left\{ 
\begin{array}{l}
\left( h_{1}^{+}\right) ^{-1}\dot{h}_{1}^{+}=\mathbf{\Omega }_{1}^{+} \\ 
\\ 
\mathbf{\dot{\Gamma}}_{1}=\Pi _{-}[\mathbf{\Gamma }_{1},\mathbf{\Omega }%
_{1}^{+}]%
\end{array}%
\right.  \label{collective ham eq. Om-Gam 0}
\end{equation}%
which, in some sense, resembles the Euler equation of motion for a rigid
body moving under dressing action in place of adjoint action.

The $\mathrm{Ad}$-invariance of $\mathsf{h}$ implies the relation%
\begin{equation*}
\mathrm{ad}_{\mathcal{L}_{\mathsf{h}}(\eta )}^{\ast }\eta =0\Longrightarrow 
\mathrm{ad}_{\Pi _{+}\mathcal{L}_{\mathsf{h}}(\eta )}^{\ast }\eta =-\mathrm{%
ad}_{\Pi _{-}\mathcal{L}_{\mathsf{h}}(\eta )}^{\ast }\eta
\end{equation*}%
that traduces in terms of the new variables $\mathbf{\Omega }_{1}\mathbf{%
,\Gamma }_{1}$ in%
\begin{equation*}
\lbrack \mathbf{\Gamma }_{1},\mathbf{\Omega }_{1}^{+}]=-[\mathbf{\Gamma }%
_{1},\mathbf{\Omega }_{1}^{-}]
\end{equation*}%
or, equivalently 
\begin{equation}
\lbrack \mathbf{\Gamma }_{1},\mathbf{\Omega }_{1}]=0  \label{Ad-invariance}
\end{equation}%
Using this result, the Hamilton equations $\left( \ref{collective ham eq.
Om-Gam 0}\right) $ now reads 
\begin{equation}
\left\{ 
\begin{array}{l}
\left( h_{1}^{+}\right) ^{-1}\dot{h}_{1}^{+}=\mathbf{\Omega }_{1}^{+} \\ 
\\ 
\mathbf{\dot{\Gamma}}_{1}=-\Pi _{-}[\mathbf{\Gamma }_{1},\mathbf{\Omega }%
_{1}^{-}]%
\end{array}%
\right.  \label{collective ham eq. Om-Gam 1}
\end{equation}

\begin{description}
\item[Remark:] \textit{Since} $\Pi _{-}^{\ast }\sigma \left( Z_{1}\right) $ 
\textit{is a character of} $\mathfrak{h}_{1}^{-}$\textit{, so is} $\Pi
_{-}^{\ast }\gamma \left( \mathbf{\Gamma }_{1}\right) =\gamma \left( \mathbf{%
\Gamma }_{1}^{+}\right) $.
\end{description}

The last Hamilton equations can still be rewritten in another form. Observe
that

\begin{equation*}
\mathbf{\dot{\Gamma}}_{1}=-\Pi _{-}\left[ \mathbf{\Gamma }_{1},\mathbf{%
\Omega }_{1}^{-}\right] =-\left[ \mathbf{\Gamma }_{1},\mathbf{\Omega }%
_{1}^{-}\right] +\Pi _{+}\left[ \mathbf{\Gamma }_{1},\mathbf{\Omega }_{1}^{-}%
\right]
\end{equation*}%
where the second term in the rhs is equivalent to%
\begin{equation*}
\Pi _{+}\left[ \mathbf{\Gamma }_{1},\mathbf{\Omega }_{1}^{-}\right] =\Pi _{+}%
\left[ \mathbf{\Gamma }_{1}^{+},\mathbf{\Omega }_{1}^{-}\right] =\gamma
^{-1}\left( \Pi _{-}^{\ast }\mathrm{ad}_{\mathbf{\Omega }_{1}^{-}}^{\ast
}\gamma \left( \mathbf{\Gamma }_{1}^{+}\right) \right)
\end{equation*}%
Here, we observe that%
\begin{equation*}
\left\langle \Pi _{-}^{\ast }\mathrm{ad}_{\mathbf{\Omega }_{1}^{-}}^{\ast
}\gamma \left( \mathbf{\Gamma }_{1}^{+}\right) ,Z\right\rangle =\left\langle 
\mathrm{ad}_{\mathbf{\Omega }_{1}^{-}}^{\ast }\gamma \left( \mathbf{\Gamma }%
_{1}^{+}\right) ,Z_{-}\right\rangle =\left\langle ad_{\mathbf{\Omega }%
_{1}^{-}}^{\ast }\gamma \left( \mathbf{\Gamma }_{1}^{+}\right)
,Z\right\rangle
\end{equation*}%
for arbitrary $Z\in \mathfrak{h}_{1}$, so we conclude that 
\begin{equation*}
\Pi _{-}^{\ast }\mathrm{ad}_{\mathbf{\Omega }_{1}^{-}}^{\ast }\gamma \left( 
\mathbf{\Gamma }_{1}^{+}\right) =ad_{\mathbf{\Omega }_{1}^{-}}^{\ast }\gamma
\left( \mathbf{\Gamma }_{1}^{+}\right) =0
\end{equation*}%
because $\gamma \left( \mathbf{\Gamma }_{+}^{1}\right) $ is a character of $%
\mathfrak{h}_{1}^{-}$, implying that%
\begin{equation}
\Pi _{+}\left[ \mathbf{\Gamma }_{1},\mathbf{\Omega }_{1}^{-}\right] =0
\label{proyector +}
\end{equation}

Thus, finally%
\begin{equation*}
\mathbf{\dot{\Gamma}}_{1}=-\left[ \mathbf{\Gamma }_{1},\mathbf{\Omega }%
_{1}^{-}\right]
\end{equation*}%
and the Hamilton equations take the form 
\begin{equation}
\left\{ 
\begin{array}{l}
\left( h_{1}^{+}\right) ^{-1}\dot{h}_{1}^{+}=\mathbf{\Omega }_{1}^{+} \\ 
\\ 
\mathbf{\dot{\Gamma}}_{1}=-\left[ \mathbf{\Gamma }_{1},\mathbf{\Omega }%
_{1}^{-}\right]%
\end{array}%
\right.  \label{collective ham eq. Om-Gam 2}
\end{equation}

\subsection{Solving by factorization}

We now see how to solve the Hamilton equations $\left( \ref{collective ham
eq. Om-Gam 2}\right) $ by factorization, showing that the solution curves
are orbits of the action of the factor curves in $H_{1}^{+}$ and $H_{1}^{-}$
of an exponential curve in $H_{1}$.

Let us start with the last version of the Hamilton equation in terms of $%
\left( \mathbf{\Omega ,\Gamma }\right) $, namely eqs. $\left( \ref%
{collective ham eq. Om-Gam 2}\right) $. As it is well known, the second of
these equations has the solution%
\begin{equation}
\mathbf{\Gamma }_{1}\left( t\right) =\mathrm{Ad}_{g_{1}^{-}\left( t\right)
}^{1}\mathbf{\Gamma }_{1\circ }  \label{Gamma Ad}
\end{equation}%
for some initial value $\mathbf{\Gamma }_{1}\left( t_{\circ }\right) =%
\mathbf{\Gamma }_{1\circ }$, provided the curve $g_{1}^{-}\left( t\right)
\in $ $H_{1}^{-}$ solves the differential equation%
\begin{equation}
\dot{g}_{1}^{-}\left( g_{1}^{-}\right) ^{-1}=\mathbf{\Omega }_{1}^{-}
\label{Omega - g-}
\end{equation}

In addition, this brings another important consequence: the problem can be
solved by factorization. Let us show how this idea works: adding the first
equation $\left( \ref{collective ham eq. Om-Gam 2}\right) $ and $\left( \ref%
{Omega - g-}\right) $ we get 
\begin{equation*}
\left( h_{1}^{+}\right) ^{-1}\dot{h}_{1}^{+}+\dot{g}_{1}^{-}\left(
g_{1}^{-}\right) ^{-1}=\mathbf{\Omega }_{1}
\end{equation*}%
that is equivalent to

\begin{equation}
\left( h_{1}^{+}g_{1}^{-}\right) ^{-1}\frac{d}{dt}\left(
h_{1}^{+}g_{1}^{-}\right) =\mathrm{Ad}_{\left( g_{1}^{-}\right) ^{-1}}%
\mathbf{\Omega }_{1}  \label{h+h- eq}
\end{equation}

By introducing the curve 
\begin{equation*}
k_{1}\left( t\right) :=h_{1}^{+}\left( t\right) g_{1}^{-}\left( t\right)
\end{equation*}%
and the map $\Theta _{1}:H_{1}\times \mathfrak{h}_{1}\longrightarrow 
\mathfrak{h}_{1}$%
\begin{equation*}
\left( h_{1}^{+}g_{1}^{-},\mathbf{\Gamma }_{1}\right) \longmapsto \Theta
_{1}\left( k_{1},\mathbf{\Gamma }_{1}\right) =\mathcal{L}_{\mathsf{h}}\left(
\gamma \left( \mathrm{Ad}_{\left( g_{1}^{-}\right) ^{-1}}\mathbf{\Gamma }%
_{1}\right) \right) =\mathrm{Ad}_{\left( g_{1}^{-}\right) ^{-1}}\mathbf{%
\Omega }_{1}
\end{equation*}%
the equation $\left( \ref{h+h- eq}\right) $ becomes in%
\begin{equation}
k_{1}^{-1}\dot{k}_{1}=\Theta _{1}\left( k_{1},\mathbf{\Gamma }_{1}\right)
\label{k Theta}
\end{equation}

Therefore, solving the differential equation $\left( \ref{k Theta}\right) $
and decomposing the solution in its factors in $H_{1}^{+}$ and $H_{1}^{-}$,
one solves the original problem. However, for general $\Theta _{1}$ this
equation can be hardly integrated. The next result is crucial for the
success of the AKS procedure.

\begin{description}
\item[Proposition:] \textit{The vector field }$on$ $H_{1}\times \mathfrak{h}%
_{1}$ \textit{defined by the assignment }%
\begin{equation*}
\left( k_{1},\mathbf{\Gamma }_{1}\right) \longmapsto \mathcal{X}\left( k_{1},%
\mathbf{\Gamma }_{1}\right) :=\left( \Theta _{1}\left( k_{1},\mathbf{\Gamma }%
_{1}\right) ,-[\mathbf{\Gamma }_{1},\Pi _{1}^{-}\mathcal{L}_{\mathsf{h}%
}\left( \gamma \left( \mathbf{\Gamma }_{1}\right) \right) ]\right)
\end{equation*}%
\textit{\ } \textit{is in the null distribution of the differential} $\Theta
_{1\ast }$ of $\Theta _{1}$\textit{.\ }
\end{description}

\textbf{Proof: }Let us calculate it as follows%
\begin{eqnarray*}
\Theta _{1\ast }\left( \mathcal{X}\left( k_{1},\mathbf{\Gamma }_{1}\right)
\right) &=&\left. \frac{d}{dt}\mathcal{L}_{\mathsf{h}}\left( \gamma \left( 
\mathrm{Ad}_{\left( g_{1}^{-}\left( t\right) \right) ^{-1}}\left( \mathbf{%
\Gamma }_{1}\left( t\right) \right) \right) \right) \right\vert _{t=0} \\
&=&\mathcal{L}_{\mathsf{h}\ast }\left. \gamma \left( \frac{d}{dt}\left( 
\mathrm{Ad}_{\left( g_{1}^{-}\left( t\right) \right) ^{-1}}\left( \mathbf{%
\Gamma }_{1}\left( t\right) \right) \right) \right) \right\vert _{t=0}
\end{eqnarray*}%
Since 
\begin{eqnarray*}
\left. \frac{d}{dt}\left( \mathrm{Ad}_{\left( g_{1}^{-}\left( t\right)
\right) ^{-1}}\left( \mathbf{\Gamma }_{1}\left( t\right) \right) \right)
\right\vert _{t=0} &=&-\mathrm{Ad}_{\left( g_{1}^{-}\right) ^{-1}}\mathrm{ad}%
_{\dot{g}_{1}^{-}\left( g_{1}^{-}\right) ^{-1}}\mathbf{\Gamma }_{1}+\mathrm{%
Ad}_{\left( g_{1}^{-}\right) ^{-1}}\mathbf{\dot{\Gamma}}_{1} \\
&=&-\mathrm{Ad}_{\left( g_{1}^{-}\right) ^{-1}}\left[ \mathbf{\Omega }%
_{1}^{-},\mathbf{\Gamma }_{1}\right] -\mathrm{Ad}_{\left( g_{1}^{-}\right)
^{-1}}[\mathbf{\Gamma }_{1},\mathbf{\Omega }_{1}^{-}] \\
&=&0
\end{eqnarray*}%
we conclude that 
\begin{equation*}
\Theta _{1\ast }\left( \mathcal{X}\left( k_{1},\mathbf{\Gamma }_{1}\right)
\right) =0
\end{equation*}%
as stated.$\blacksquare $

\begin{description}
\item[Corollary:] \textit{The map }$\Theta ^{1}:H_{1}\times \mathfrak{h}%
_{1}\longrightarrow \mathfrak{h}_{1}$ \textit{is constant along the solution
curves of the system of Hamilton equations}%
\begin{equation*}
\left\{ 
\begin{array}{l}
k_{1}^{-1}\dot{k}_{1}=\Theta _{1}\left( k_{1},\mathbf{\Gamma }_{1}\right) \\ 
\\ 
\mathbf{\dot{\Gamma}}_{1}=-\left[ \mathbf{\Gamma }_{1},\mathbf{\Omega }%
_{1}^{-}\right]%
\end{array}%
\right.
\end{equation*}
\end{description}

It makes the differential equation on $H_{1}$ $\left( \ref{k Theta}\right) $
easily integrable: because $\Theta _{1}\left( k_{1},\mathbf{\Gamma }%
_{1}\right) $ is a constant of motion $\Theta _{1\circ }\in \mathfrak{h}_{1}$%
, it has the exponential solution 
\begin{equation*}
k\left( t\right) =e^{t\Theta _{1\circ }}
\end{equation*}%
The decomposition of this solution $k\left( t\right) $\ in its factor on $%
H_{1}^{+}$ and $H_{1}^{-}$, as 
\begin{equation*}
e^{t\Theta _{1\circ }}=h_{1}^{+}\left( t\right) g_{1}^{-}\left( t\right)
\end{equation*}%
gives the full solution of the original problem. In fact, 
\begin{eqnarray*}
\mathbf{\Omega }_{1}^{+} &=&\left( h_{1}^{+}\right) ^{-1}\dot{h}_{1}^{+} \\
&& \\
\mathbf{\Omega }_{1}^{-} &=&\dot{g}_{1}^{-}\left( g_{1}^{-}\right) ^{-1}
\end{eqnarray*}%
and%
\begin{equation}
\mathbf{\Gamma }_{1}\left( t\right) =\mathrm{Ad}_{g_{1}^{-}\left( t\right)
}^{1}\mathbf{\Gamma }_{1\circ }  \label{solution Gamma 1}
\end{equation}%
for some initial value $\mathbf{\Gamma }_{1}\left( t_{\circ }\right) =%
\mathbf{\Gamma }_{1\circ }$.

\subsection{The nested equation of motion \label{nested}}

One may think of phase spaces on semidirect product Lie groups as way of
packing many variables in a well suited fashion in order to formulate a
complicated problem in a simpler way. It becomes evident if we consider the
Hamilton equation on $H_{1}=H_{0}\circledS \mathfrak{h}_{0}$ and we express
it in terms of the $H_{0}$ and $\mathfrak{h}_{0}$ variables. In doing so, we
consider now the equations $\left( \ref{collective ham eq. on N}\right) $ on 
$\mathcal{N}\left( h_{1}^{-},Z_{1}^{-}\right) $, in terms of the variables $%
\left( \mathbf{\Omega }_{1},\mathbf{\Gamma }_{1}\right) $ as in eqs. $\left( %
\ref{collective ham eq. Om-Gam 0},\ref{collective ham eq. Om-Gam 1},\ref%
{collective ham eq. Om-Gam 2}\right) $, and write them in terms the
variables in $H_{0},\mathfrak{h}_{0}$. Let us first to establish some
notational convention regarding the elements of $H_{1}$ as objects in $%
H_{0}\times \mathfrak{h}_{0}$:%
\begin{equation*}
\begin{array}{l}
\left( h_{1},Z_{1}\right) :=\left( \left( h_{0},Z_{0}\right) ,\left(
X_{0},Y_{0}\right) \right) \\ 
\\ 
\gamma ^{-1}\left( \sigma \left( Z_{1}\right) \right) :=\left( \gamma
^{-1}\left( \sigma \left( Y_{0}\right) \right) ,\gamma ^{-1}\left( \sigma
\left( X_{0}\right) \right) \right) \\ 
\\ 
\mathcal{L}_{\mathsf{h}}\left( \sigma \left( Z_{1}\right) \right) :=\left( 
\mathcal{L}_{\mathsf{h}}^{\prime }\left( \sigma \left( Z_{1}\right) \right) ,%
\mathcal{L}_{\mathsf{h}}^{\prime \prime }\left( \sigma \left( Z_{1}\right)
\right) \right)%
\end{array}%
\end{equation*}%
and, having in mind the expression for the corresponding adjoint action
given in $\left( \ref{Hxh-5},\ref{Hxh-6}\right) $, we have that 
\begin{eqnarray*}
\mathbf{\Omega }_{1} &=&\left( \mathrm{Ad}_{h_{0}^{-}}^{0}\mathcal{L}_{%
\mathsf{h}}^{\prime }\left( \sigma \left( Z_{1}\right) \right) \right. ,\, \\
&&,\left. \left[ \mathrm{Ad}_{h_{0}^{-}}^{0}Z_{0}^{-},\mathrm{Ad}%
_{h_{0}^{-}}^{0}\mathcal{L}_{\mathsf{h}}^{\prime }\left( \sigma \left(
Z_{1}\right) \right) \right] +\mathrm{Ad}_{h_{0}^{-}}^{0}\mathcal{L}_{%
\mathsf{h}}^{\prime \prime }\left( \sigma \left( Z_{1}\right) \right) \right)
\\
&& \\
\mathbf{\Gamma }_{1} &=&\left( \mathrm{Ad}_{h_{0}^{-}}^{0}\gamma ^{-1}\left(
\sigma \left( Y_{0}\right) \right) \right. ,\, \\
&&,\left. \left[ \mathrm{Ad}_{h_{0}^{-}}^{0}Z_{0}^{-},\mathrm{Ad}%
_{h_{0}^{-}}^{0}\gamma ^{-1}\left( \sigma \left( Y_{0}\right) \right) \right]
+\mathrm{Ad}_{h_{0}^{-}}^{0}\gamma ^{-1}\left( \sigma \left( X_{0}\right)
\right) \right)
\end{eqnarray*}

By analogy, we introduce the variables

\begin{eqnarray}
\mathbf{\tilde{\Omega}}_{0} &\mathbf{=}&\mathrm{Ad}_{h_{0}^{-}}^{0}\mathcal{L%
}_{\mathsf{h}}^{\prime }\left( \sigma \left( X_{0},Y_{0}\right) \right) 
\notag \\
\mathbf{N}_{0} &\mathbf{=}&\mathrm{Ad}_{h_{0}^{-}}^{0}\mathcal{L}_{\mathsf{h}%
}^{\prime \prime }\left( \sigma \left( X_{0},Y_{0}\right) \right)  \notag \\
\mathbf{\tilde{\Gamma}}_{0} &\mathbf{=}&\mathrm{Ad}_{h_{0}^{-}}^{0}\gamma
^{-1}\left( \sigma \left( Y_{0}\right) \right)  \label{Omega 0 - Gamma 0} \\
\mathbf{M}_{0} &\mathbf{=}&\mathrm{Ad}_{h_{0}^{-}}^{0}\gamma ^{-1}\left(
\sigma \left( X_{0}\right) \right)  \notag \\
\mathbf{R}_{0}^{-} &=&\mathrm{Ad}_{h_{0}^{-}}^{0}Z_{0}^{-}  \notag
\end{eqnarray}%
so, the above expressions for $\left( \mathbf{\Omega }_{1},\mathbf{\Gamma }%
_{1}\right) $ become 
\begin{eqnarray}
\mathbf{\Omega }_{1} &=&\left( \mathbf{\tilde{\Omega}}_{0},\left[ \mathbf{R}%
_{0}^{-},\mathbf{\tilde{\Omega}}_{0}\right] +\mathbf{N}_{0}\right)  \notag \\
&&  \label{Kirchoff nested eqs 0} \\
\mathbf{\Gamma }_{1} &=&\left( \mathbf{\tilde{\Gamma}}_{0},\left[ \mathbf{R}%
_{0}^{-},\mathbf{\tilde{\Gamma}}_{0}\right] +\mathbf{M}_{0}\right)  \notag
\end{eqnarray}

Also, the (translated) evolution vector field reads 
\begin{equation*}
\left( h_{1}^{+}\right) ^{-1}\left( t\right) \dot{h}_{1}^{+}\left( t\right)
=\left( \left( h_{0}^{+}\right) ^{-1}\left( t\right) \dot{h}_{0}^{+}\left(
t\right) ,\left[ \left( h_{0}^{+}\left( t\right) \right) ^{-1}\dot{h}%
_{0}^{+}\left( t\right) ,Z_{0}^{+}\left( t\right) \right] +\dot{Z}%
_{0}^{+}\left( t\right) \right)
\end{equation*}%
and%
\begin{eqnarray*}
&&[\mathbf{\Gamma }_{1},\mathbf{\Omega }_{1}^{+}] \\
&=&\left( \left[ \mathbf{\tilde{\Gamma}}_{0},\mathbf{\tilde{\Omega}}_{0}^{+}%
\right] \right. , \\
&&\quad ,\left. \left[ \mathbf{\tilde{\Gamma}}_{0},\Pi _{+}\left[ \mathbf{R}%
_{0}^{-},\mathbf{\tilde{\Omega}}_{0}\right] \right] +\left[ \mathbf{\tilde{%
\Gamma}}_{0},\mathbf{N}_{0}^{+}\right] +\left[ \left[ \mathbf{R}_{0}^{-},%
\mathbf{\tilde{\Gamma}}_{0}\right] ,\mathbf{\tilde{\Omega}}_{0}^{+}\right] +%
\left[ \mathbf{M}_{0},\mathbf{\tilde{\Omega}}_{0}^{+}\right] \right)
\end{eqnarray*}

Therefore, the equations of motion $\left( \ref{collective ham eq. Om-Gam 0}%
\right) $ for the $H_{0}^{\pm },\mathfrak{h}_{0}^{\pm }$ coordinates are 
\begin{equation}
\left\{ 
\begin{array}{l}
\left( h_{0}^{+}\right) ^{-1}\left( t\right) \dot{h}_{0}^{+}\left( t\right) =%
\mathbf{\tilde{\Omega}}_{0}^{+} \\ 
\\ 
\dot{Z}_{0}^{+}\left( t\right) =-\left[ \mathbf{\tilde{\Omega}}%
_{0}^{+},Z_{0}^{+}\right] +\left[ \mathbf{R}_{-}^{0},\mathbf{\tilde{\Omega}}%
_{0}\right] +\mathbf{N}_{0} \\ 
\\ 
\dfrac{d}{dt}\mathbf{\tilde{\Gamma}}_{0}=\Pi _{-}\left[ \mathbf{\tilde{\Gamma%
}}_{0},\mathbf{\tilde{\Omega}}_{0}^{+}\right] \\ 
\\ 
\mathbf{\dot{M}}_{0}=\Pi _{-}\left[ \mathbf{\tilde{\Gamma}}_{0},\mathbf{N}%
_{0}^{+}\right] +\Pi _{-}\left[ \mathbf{M}_{0},\mathbf{\tilde{\Omega}}%
_{0}^{+}\right] +\left[ \Pi _{-}\left[ \mathbf{\tilde{\Gamma}}_{0},\mathbf{%
\tilde{\Omega}}_{0}^{+}\right] ,\mathbf{R}_{0}^{-}\right] \\ 
\qquad \quad +\Pi _{-}\left[ \mathbf{\tilde{\Gamma}}_{0},\Pi _{+}\left[ 
\mathbf{R}_{0}^{-},\mathbf{\tilde{\Omega}}_{0}\right] \right] +\Pi _{-}\left[
\left[ \mathbf{R}_{0}^{-},\mathbf{\tilde{\Gamma}}_{0}\right] ,\mathbf{\tilde{%
\Omega}}_{0}^{+}\right]%
\end{array}%
\right.  \label{Kirchoff nested eqs 1}
\end{equation}%
They become simpler and more familiar for $\mathbf{R}_{0}^{-}=0$%
\begin{equation}
\left\{ 
\begin{array}{l}
\left( h_{0}^{+}\right) ^{-1}\dfrac{d}{dt}h_{0}^{+}=\mathbf{\tilde{\Omega}}%
_{0}^{+} \\ 
\\ 
\dfrac{d}{dt}Z_{0}^{+}=-\left[ \mathbf{\tilde{\Omega}}_{0}^{+},Z_{0}^{+}%
\right] +\mathbf{N}_{0} \\ 
\\ 
\dfrac{d}{dt}\mathbf{\tilde{\Gamma}}_{0}=\Pi _{-}\left[ \mathbf{\tilde{\Gamma%
}}_{0},\mathbf{\tilde{\Omega}}_{0}^{+}\right] \\ 
\\ 
\dfrac{d}{dt}\mathbf{M}_{0}=\Pi _{-}\left[ \mathbf{\tilde{\Gamma}}_{0},%
\mathbf{N}_{0}^{+}\right] +\Pi _{-}\left[ \mathbf{M}_{0},\mathbf{\tilde{%
\Omega}}_{0}^{+}\right]%
\end{array}%
\right.  \label{Kirchoff nested eqs 2}
\end{equation}

\begin{description}
\item[Remark:] Observe that in the last two equations we can replace 
\begin{eqnarray*}
\Pi _{-}\left[ \mathbf{\tilde{\Gamma}}_{0},\mathbf{\tilde{\Omega}}_{0}^{+}%
\right] &=&\Pi _{-}\left[ \mathbf{\tilde{\Gamma}}_{0}^{-},\mathbf{\tilde{%
\Omega}}_{0}^{+}\right] =\left( \mathbf{\tilde{\Gamma}}_{0}^{-}\right) ^{%
\mathbf{\tilde{\Omega}}_{0}^{+}} \\
&& \\
\Pi _{-}\left[ \mathbf{M}_{0},\mathbf{\tilde{\Omega}}_{0}^{+}\right] &=&\Pi
_{-}\left[ \mathbf{M}_{0}^{-},\mathbf{\tilde{\Omega}}_{0}^{+}\right] =\left( 
\mathbf{M}_{0}^{-}\right) ^{\mathbf{\tilde{\Omega}}_{0}^{+}}
\end{eqnarray*}%
where $\left( Y_{0}^{-}\right) ^{X_{0}^{+}}$ means the dressing action of $%
\mathfrak{h}_{0}^{+}$ on $\mathfrak{h}_{0}^{-}$ (see eq. $\left( \ref{dlg-2a}%
\right) $ and ref. \cite{Lu-We}). Then, these equations turn into%
\begin{equation*}
\left\{ 
\begin{array}{l}
\dfrac{d}{dt}\mathbf{\tilde{\Gamma}}_{0}^{-}=\left( \mathbf{\tilde{\Gamma}}%
_{0}^{-}\right) ^{\mathbf{\tilde{\Omega}}_{0}^{+}} \\ 
\\ 
\dfrac{d}{dt}\mathbf{M}_{0}^{-}=\left( \mathbf{\tilde{\Gamma}}%
_{0}^{-}\right) ^{\mathbf{N}_{0}^{+}}+\left( \mathbf{M}_{0}^{-}\right) ^{%
\mathbf{\tilde{\Omega}}_{0}^{+}}%
\end{array}%
\right.
\end{equation*}%
that resembles the Euler-Poisson equations of the heavy rigid body. In fact,
they are 
\begin{equation*}
\left\{ 
\begin{array}{l}
\dot{\Gamma}=\left[ \Gamma ,\Omega \right] \\ 
\\ 
\dot{M}=\left[ M,\Omega \right] +\left[ \Gamma ,N\right]%
\end{array}%
\right.
\end{equation*}%
where $\Omega $ is the angular velocity, $M$ is the angular momenta, $\Gamma 
$ is the gravitational force as seen from the moving frame, and $N$ is the
vector joining the center of mass with a fixed point. In these equations the
evolution is generated by the adjoint action of the Lie algebra $\mathfrak{so%
}_{3}$ on itself, while in our case is generated by the dressing action of $%
\mathfrak{h}_{0}^{+}$ on $\mathfrak{h}_{0}^{-}$.
\end{description}

In order to solve the equations $\left( \ref{Kirchoff nested eqs 1}\right) $
we just need to factorize the exponential curve $k_{1}\left( t\right) $ on $%
H_{1}$%
\begin{equation*}
k_{1}\left( t\right) =e^{t\Theta _{1}}=h_{1}^{+}\left( t\right)
g_{1}^{-}\left( t\right)
\end{equation*}%
with 
\begin{equation*}
\mathbf{\Omega }_{1}\left( t\right) =\mathrm{Ad}_{g_{1}^{-}\left( t\right)
}^{1}\Theta _{1}
\end{equation*}%
These factors decompose on $H_{0}^{\pm }$ and $\mathfrak{h}_{0}^{\pm }$ as 
\begin{equation*}
\left\{ 
\begin{array}{l}
h_{1}^{+}\left( t\right) =\left( h_{0}^{+}\left( t\right) ,Z_{0}^{+}\left(
t\right) \right) \\ 
g_{1}^{-}\left( t\right) =\left( g_{0}^{-}\left( t\right) ,Z_{0}^{-}\left(
t\right) \right)%
\end{array}%
\right.
\end{equation*}%
so that from the solution $\left( \ref{solution Gamma 1}\right) $ 
\begin{equation*}
\mathbf{\Gamma }_{1}\left( t\right) =\mathrm{Ad}_{g_{1}^{-}\left( t\right)
}^{1}\mathbf{\Gamma }_{1\circ }=\left( \mathrm{Ad}_{g_{0}^{-}\left( t\right)
}^{0}\mathbf{\tilde{\Gamma}}_{0\circ }^{\prime },\mathrm{Ad}%
_{g_{0}^{-}\left( t\right) }^{0}\left( \left[ Z_{0}^{-}\left( t\right) ,%
\mathbf{\tilde{\Gamma}}_{0\circ }^{\prime }\right] +\mathbf{\tilde{\Gamma}}%
_{0\circ }^{\prime \prime }\right) \right)
\end{equation*}%
we get%
\begin{equation*}
\mathbf{\Gamma }_{1}\left( t\right) =\left( \mathbf{\tilde{\Gamma}}%
_{0}\left( t\right) ,\left[ \mathbf{R}_{0}^{-},\mathbf{\tilde{\Gamma}}%
_{0}\left( t\right) \right] +\mathbf{M}_{0}\left( t\right) \right)
\end{equation*}%
that implies%
\begin{eqnarray}
\mathbf{\tilde{\Gamma}}_{0}\left( t\right) &=&\mathrm{Ad}_{g_{0}^{-}\left(
t\right) }^{0}\mathbf{\tilde{\Gamma}}_{0\circ }^{\prime }  \notag \\
&&  \label{Kirchoff nested eqs 3} \\
\mathbf{M}_{0}\left( t\right) &=&\left[ \mathrm{Ad}_{g_{0}^{-}\left(
t\right) }^{0}Z_{0}^{-}\left( t\right) -\mathbf{R}_{0}^{-},\mathrm{Ad}%
_{g_{0}^{-}\left( t\right) }^{0}\mathbf{\tilde{\Gamma}}_{0\circ }^{\prime }%
\right] +\mathrm{Ad}_{g_{0}^{-}\left( t\right) }^{0}\mathbf{\tilde{\Gamma}}%
_{0\circ }^{\prime \prime }  \notag
\end{eqnarray}%
for some initial value $\mathbf{\Gamma }_{1\circ }=\left( \mathbf{\tilde{%
\Gamma}}_{0\circ }^{\prime },\mathbf{\tilde{\Gamma}}_{0\circ }^{\prime
\prime }\right) $.

\section{Tower of integrable systems}

The above construction can be made recursive for appropriated hamiltonians,
giving rise thus to a tower of integrable system on semidirect products, all
of them solvable by factorization.

First, let us suppose that $G$ is a \emph{semisimple Lie group}, $\mathfrak{g%
}$ is its Lie algebra, and $\left( \cdot ,\cdot \right) _{\mathfrak{g}}$ its
Killing form. For every $m=1,2,\dots $ we define $H_{m}:=H_{m-1}\circledS 
\mathfrak{h}_{m-1}$, and we identify $H_{0}=G$, $\mathfrak{h}_{0}=\mathfrak{g%
}$ and $\mathsf{k}_{0}=\left( \cdot ,\cdot \right) _{\mathfrak{g}}$. Then we
have:

\begin{description}
\item[Lemma:] \textit{Let }$\mathfrak{h}_{m-1}$ \textit{be a Lie algebra
equipped with a nondegenerate }$\mathit{\mathrm{Ad}}^{m-1}$\textit{%
-invariant bilinear form }$\mathsf{k}_{m-1}:\mathfrak{h}_{m-1}\otimes 
\mathfrak{h}_{m-1}\longrightarrow \mathbb{K}$. \textit{On} $\mathfrak{h}%
_{m}:=\mathfrak{h}_{m-1}\circledS \mathfrak{h}_{m-1}$ \textit{the bilinear
form} $\mathsf{k}_{m}:\mathfrak{h}_{m}\times \mathfrak{h}_{m}\rightarrow 
\mathbb{K}$ \textit{given by} 
\begin{equation*}
\mathsf{k}_{m}\left( \left( X,Y\right) ,\left( X^{\prime },Y^{\prime
}\right) \right) :=\frac{1}{2}\left( \mathsf{k}_{m-1}\left( X,Y^{\prime
}\right) +\mathsf{k}_{m-1}\left( Y,X^{\prime }\right) \right)
\end{equation*}%
$\forall $\textit{\ }$\left( X,Y\right) ,\left( X^{\prime },Y^{\prime
}\right) \in \mathfrak{h}_{m}$\textit{, is nondegenerate and }$\mathrm{Ad}%
^{m}$\textit{-invariant}.
\end{description}

\textbf{Proof: }It follows by direct substitution of 
\begin{equation*}
\mathrm{Ad}_{\left( b,Z\right) }^{m}\left( X,Y\right) =\left( \mathrm{Ad}%
_{b}^{m-1}X\,,\mathrm{Ad}_{b}^{m-1}\left( \left[ Z,X\right] _{m-1}+Y\right)
\right)
\end{equation*}%
for $\left( b,Z\right) \in H_{m}$ and $\left( X,Y\right) ,\left( X^{\prime
},Y^{\prime }\right) \in \mathfrak{h}_{m}$, in 
\begin{equation*}
\mathsf{k}_{m}\left( \mathrm{Ad}_{\left( b,Z\right) }^{m}\left( X,Y\right) ,%
\mathrm{Ad}_{\left( b,Z\right) }^{m}\left( X^{\prime },Y^{\prime }\right)
\right)
\end{equation*}%
and using the $\mathit{\mathrm{Ad}}^{m-1}$-invariance of the bilinear form%
\textit{\ }$\mathsf{k}_{m-1}$, to reach 
\begin{equation*}
\mathsf{k}_{m}\left( \mathrm{Ad}_{\left( b,Z\right) }^{m}\left( X,Y\right) ,%
\mathrm{Ad}_{\left( b,Z\right) }^{m}\left( X^{\prime },Y^{\prime }\right)
\right) =\mathsf{k}_{m}\left( \left( X_{1},Y_{1}\right) ,\left( X^{\prime
},Y^{\prime }\right) \right)
\end{equation*}%
as expected.$\blacksquare $

As we saw above, eqs. $\left( \ref{factor H1},\ref{factorization semidirect
2}\right) $, the factorization 
\begin{equation*}
\left\{ 
\begin{array}{l}
H_{m}=H_{m}^{+}H_{m}^{-} \\ 
\mathfrak{h}_{m}=\mathfrak{h}_{m}^{+}\oplus \mathfrak{h}_{m}^{-}%
\end{array}%
\right.
\end{equation*}%
implies that $H_{m+1}$ factorizes according to 
\begin{equation*}
H_{m+1}=H_{m+1}^{+}H_{m+1}^{-}=\left( H_{m}^{+}\circledS \mathfrak{h}%
_{m}^{+}\right) \cdot \left( H_{m}^{-}\circledS \mathfrak{h}_{m}^{-}\right)
\end{equation*}%
Explicitly, it means that $\left( h,Y\right) \in H_{m+1}=H_{m}\circledS 
\mathfrak{h}_{m}$ admits the factorization%
\begin{equation*}
\left( h,Y\right) =\left( h_{+},X_{+}\right) \cdot \left( h_{-},X_{-}\right)
\end{equation*}%
with 
\begin{equation*}
\left\{ 
\begin{array}{l}
h=h_{+}h_{-} \\ 
X_{+}=\Pi _{+}\left( \mathrm{Ad}_{h_{-}}^{m}Y\right) \\ 
X_{-}=\mathrm{Ad}_{h_{-}^{-1}}^{m}\Pi _{-}\mathrm{Ad}_{h_{-}}^{m}Y%
\end{array}%
\right.
\end{equation*}

The chain of semidirect product of Lie groups produces a chain of semidirect
sum of Lie algebras. Each element factorizes as%
\begin{equation*}
\mathfrak{h}_{m+1}=\mathfrak{h}_{m+1}^{+}\circledS \mathfrak{h}_{m+1}^{-}
\end{equation*}%
such that $\left( X,Y\right) \in \mathfrak{h}_{m+1}=\mathfrak{h}%
_{m}\circledS \mathfrak{h}_{m}$, $\left( X_{\pm },Y_{\pm }\right) \in 
\mathfrak{h}_{m+1}^{\pm }=\mathfrak{h}_{m}^{\pm }\circledS \mathfrak{h}%
_{m}^{\pm }$, factorizes as $\ $ 
\begin{equation*}
\left( X,Y\right) =\left( X_{+},Y_{+}\right) \oplus \left( X_{-},Y_{-}\right)
\end{equation*}

Let us choose the hamiltonian on $\mathcal{N}_{m+1}\left(
h_{m}^{-},Z_{m}^{-}\right) \subset H_{m+1}$ as 
\begin{equation}
\mathcal{H}_{m+1}(h_{m},Z_{m}):=\mathsf{h}^{\left( m\right) }\left( \Phi
^{\left( m+1\right) }(h_{m},Z_{m})\right)  \label{cuadratic ham}
\end{equation}%
for the momentum map%
\begin{equation*}
\Phi ^{\left( m+1\right) }(h_{m},Z_{m})=\gamma \left( \mathrm{Ad}%
_{h_{m}}^{m}\gamma ^{-1}\left( \sigma \left( Z_{m}\right) \right) \right) =%
\mathrm{Ad}_{h_{m}^{-1}}^{\left( m\right) \ast }\sigma \left( Z_{m}\right)
\end{equation*}%
and $\mathsf{h}^{\left( m\right) }:\mathfrak{h}_{m}^{\ast }\longrightarrow 
\mathbb{R}$ 
\begin{equation*}
\mathsf{h}^{\left( m\right) }\left( \eta \right) =\dfrac{1}{2}\mathsf{k}%
_{m}\left( \gamma ^{-1}\left( \eta \right) ,\gamma ^{-1}\left( \eta \right)
\right)
\end{equation*}%
Then, its Legendre transform $\mathcal{L}_{\mathsf{h}^{\left( m\right) }}:%
\mathfrak{h}_{m}^{\ast }\longrightarrow \mathfrak{h}_{m}$ 
\begin{equation*}
\mathsf{k}_{m}\left( \mathcal{L}_{\mathsf{h}^{\left( m\right) }}(\eta
_{1}),Y_{1}\right) =\left. \frac{d}{dt}\mathsf{h}_{m}\left( \eta
_{m}+t\gamma \left( Y_{m}\right) \right) \right\vert _{t=0}
\end{equation*}%
is%
\begin{equation*}
\mathcal{L}_{\mathsf{h}^{\left( m\right) }}(\eta _{m})=\gamma ^{-1}\left(
\eta _{m}\right) =\left( \gamma ^{-1}(\eta _{m-1}^{\prime }),\gamma
^{-1}(\eta _{m-1})\right)
\end{equation*}

In terms of the variables $\left( \mathbf{\Omega ,\Gamma }\right) $, see $%
\left( \ref{Omega-Gamma}\right) $, we have%
\begin{equation*}
\mathbf{\Omega }_{m}\mathbf{=}\mathrm{Ad}_{h_{m}^{-}}^{m}\gamma ^{-1}\left(
\sigma \left( Z_{m}\right) \right) =\mathbf{\Gamma }_{m}
\end{equation*}%
and the collective Hamilton equations $\left( \ref{collective ham eq. on N}%
\right) $ are 
\begin{equation}
\left\{ 
\begin{array}{l}
\left( h_{m}^{+}\right) ^{-1}\dot{h}_{m}^{+}=\mathbf{\Omega }_{m}^{+} \\ 
\\ 
\mathbf{\dot{\Omega}}_{m}=\Pi _{-}[\mathbf{\Omega }_{m}^{-},\mathbf{\Omega }%
_{m}^{+}]%
\end{array}%
\right.  \label{torre m-1}
\end{equation}%
To write down the dynamics in terms of the coordinates of $H_{m-1}^{+}$ and $%
\mathfrak{h}_{m-1}^{+}$, setting 
\begin{equation*}
\left( h_{m},Z_{m}\right) =\left( \left( h_{m-1},Z_{m-1}\right) ,\left(
X_{m-1},Y_{m-1}\right) \right)
\end{equation*}%
we introduce the variables 
\begin{eqnarray*}
\mathbf{\tilde{\Omega}}_{m-1} &=&\mathbf{\tilde{\Gamma}}_{m-1}=\mathrm{Ad}%
_{h_{m-1}^{-}}^{m-1}\gamma ^{-1}(\sigma \left( X_{m-1}\right) ) \\
\mathbf{N}_{m-1} &\mathbf{=}&\mathrm{Ad}_{h_{m-1}^{-}}^{m-1}\gamma
^{-1}(\sigma \left( Y_{m-1}\right) ) \\
\mathbf{M}_{m-1} &\mathbf{=}&\mathrm{Ad}_{h_{m-1}^{-}}^{m-1}\gamma
^{-1}\left( \sigma \left( Y_{m-1}\right) \right) \\
\mathbf{R}_{m-1}^{-} &=&\mathrm{Ad}_{h_{m-1}^{-}}^{m-1}Z_{m-1}^{-}
\end{eqnarray*}%
such that $\mathbf{\Omega }_{m}$ and $\mathbf{\Gamma }_{m}$ are now%
\begin{eqnarray*}
\mathbf{\Omega }_{m} &=&\left( \mathbf{\tilde{\Omega}}_{m-1},\left[ \mathbf{R%
}_{m-1}^{-},\mathbf{\tilde{\Omega}}_{m-1}\right] +\mathbf{N}_{m-1}\right) \\
&& \\
\mathbf{\Gamma }_{m} &=&\left( \mathbf{\tilde{\Gamma}}_{m-1},\left[ \mathbf{R%
}_{m-1}^{-},\mathbf{\tilde{\Gamma}}_{m-1}\right] +\mathbf{M}_{m-1}\right)
\end{eqnarray*}%
and the corresponding equations of motion $\left( \ref{Kirchoff nested eqs 1}%
\right) $ are 
\begin{equation*}
\left\{ 
\begin{array}{l}
\left( h_{m-1}^{+}\right) ^{-1}\left( t\right) \dot{h}_{m-1}^{+}\left(
t\right) =\mathbf{\tilde{\Omega}}_{m-1}^{+} \\ 
\\ 
\dot{Z}_{m-1}^{+}\left( t\right) =-\left[ \mathbf{\tilde{\Omega}}%
_{m-1}^{+},Z_{m-1}^{+}\right] +\left[ \mathbf{R}_{m-1}^{-},\mathbf{\tilde{%
\Omega}}_{m-1}\right] +\mathbf{N}_{m-1} \\ 
\\ 
\dfrac{d}{dt}\mathbf{\tilde{\Gamma}}_{m-1}=\Pi _{-}\left[ \mathbf{\tilde{%
\Gamma}}_{m-1},\mathbf{\tilde{\Omega}}_{m-1}^{+}\right] \\ 
\\ 
\mathbf{\dot{M}}_{m-1}=\Pi _{-}\left[ \mathbf{\tilde{\Gamma}}_{m-1},\mathbf{N%
}_{m-1}^{+}\right] +\Pi _{-}\left[ \mathbf{M}_{m-1},\mathbf{\tilde{\Omega}}%
_{m-1}^{+}\right] \\ 
\qquad \quad \quad +\left[ \Pi _{-}\left[ \mathbf{\tilde{\Gamma}}_{m-1},%
\mathbf{\tilde{\Omega}}_{m-1}^{+}\right] ,\mathbf{R}_{m-1}^{-}\right] \\ 
\qquad \quad \quad +\Pi _{-}\left[ \mathbf{\tilde{\Gamma}}_{m-1},\Pi _{+}%
\left[ \mathbf{R}_{m-1}^{-},\mathbf{\tilde{\Omega}}_{m-1}\right] \right] \\ 
\qquad \quad \quad +\Pi _{-}\left[ \left[ \mathbf{R}_{m-1}^{-},\mathbf{%
\tilde{\Gamma}}_{m-1}\right] ,\mathbf{\tilde{\Omega}}_{m-1}^{+}\right]%
\end{array}%
\right.
\end{equation*}

A simpler set of equations is obtained by setting the constant point $%
Z_{m-1}^{-}=0$, which implies that $\mathbf{R}_{m-1}^{-}=\mathrm{Ad}%
_{h_{m-1}^{-}}^{m-1}Z_{m-1}^{-}=0$, 
\begin{equation*}
\left\{ 
\begin{array}{l}
\left( h_{m-1}^{+}\right) ^{-1}\left( t\right) \dot{h}_{m-1}^{+}\left(
t\right) =\mathbf{\tilde{\Omega}}_{m-1}^{+} \\ 
\\ 
\dot{Z}_{m-1}^{+}\left( t\right) =-\left[ \mathbf{\tilde{\Omega}}%
_{m-1}^{+},Z_{m-1}^{+}\right] +\mathbf{N}_{m-1} \\ 
\\ 
\dfrac{d}{dt}\mathbf{\tilde{\Gamma}}_{m-1}=\Pi _{-}\left[ \mathbf{\tilde{%
\Gamma}}_{m-1},\mathbf{\tilde{\Omega}}_{m-1}^{+}\right] \\ 
\\ 
\mathbf{\dot{M}}_{m-1}=\Pi _{-}\left[ \mathbf{\tilde{\Gamma}}_{m-1},\mathbf{N%
}_{m-1}^{+}\right] +\Pi _{-}\left[ \mathbf{M}_{m-1},\mathbf{\tilde{\Omega}}%
_{m-1}^{+}\right]%
\end{array}%
\right.
\end{equation*}%
besides the equations $d\mathbf{\tilde{\Gamma}}_{m-1}^{+}/dt=d\mathbf{M}%
_{m-1}^{+}/dt=0$.

Thus, the dynamical system $\left( \ref{torre m-1}\right) $ replicates one
level down in the tower by projecting$\ $on the second component of the
semidirect product, and all of them are solvable by factorization.

\section{Example: $SL\left( 2,\mathbb{C}\right) $}

\subsection{Iwasawa decomposition of $SL(2,\mathbb{C})$}

Let us consider the Lie algebra $\mathfrak{sl}_{2}(\mathbb{C})$,with the
Iwasawa decomposition of $\mathfrak{h}_{0}=\mathfrak{sl}_{2}(\mathbb{C})^{%
\mathbb{R}}$ by taking $\mathfrak{h}_{0}^{+}=\mathfrak{su}_{2}$ and $%
\mathfrak{h}_{0}^{-}=\mathfrak{b}$,%
\begin{equation*}
\mathfrak{sl}_{2}(\mathbb{C})^{\mathbb{R}}=\mathfrak{su}_{2}\oplus \mathfrak{%
b}
\end{equation*}%
where $\mathfrak{su}_{2}$ is the real subalgebra of $\mathfrak{sl}_{2}(%
\mathbb{C})$ of antihermitean matrices, and $\mathfrak{b}$ is the subalgebra
of upper triangular matrices with real diagonal and null trace. For $%
\mathfrak{su}_{2}$ we take the basis 
\begin{equation}
\begin{array}{ccccc}
X_{1}=\left( 
\begin{array}{cc}
0 & i \\ 
i & 0%
\end{array}%
\right) &  & X_{2}=\left( 
\begin{array}{cc}
0 & 1 \\ 
-1 & 0%
\end{array}%
\right) &  & X_{3}=\left( 
\begin{array}{cc}
i & 0 \\ 
0 & -i%
\end{array}%
\right)%
\end{array}
\label{su2 basis}
\end{equation}%
and in $\mathfrak{b}$ this one%
\begin{equation}
\begin{array}{ccccc}
E=\left( 
\begin{array}{cc}
0 & 1 \\ 
0 & 0%
\end{array}%
\right) ~, &  & iE=\left( 
\begin{array}{cc}
0 & i \\ 
0 & 0%
\end{array}%
\right) ~, &  & H=\left( 
\begin{array}{cc}
1 & 0 \\ 
0 & -1%
\end{array}%
\right)%
\end{array}
\label{b basis}
\end{equation}%
We also introduce the dual basis $\left\{ \mathbf{x}_{1},\mathbf{x}_{2},%
\mathbf{x}_{3}\right\} \subset \mathfrak{su}_{2}^{\ast }$ and $\left\{ 
\mathbf{e,\tilde{e},h}\right\} \subset \mathfrak{b}^{\ast }$. This
decomposition translate to the group giving $SL\left( 2,\mathbb{C}\right)
=SU\left( 2\right) \times B$, where now $B$ is the group of $2\times 2$
upper triangular matrices with real diagonal and determinant $1$. So, in
order to fit the previous notation we identify $\mathfrak{h}_{0}=\mathfrak{sl%
}_{2}(\mathbb{C})$, $\mathfrak{h}_{0}^{+}=\mathfrak{su}_{2}$, $\mathfrak{h}%
_{0}^{-}=\mathfrak{b}$, and $H_{0}^{+}=SU\left( 2\right) $ and $H_{0}^{-}=B$.

The Lie algebra of $\mathfrak{sl}_{2}(\mathbb{C})$ in this basis is 
\begin{equation}
\begin{array}{lllll}
\left[ X_{1},X_{2}\right] =-2X_{3} &  & \left[ X_{3},X_{1}\right] =-2X_{2} & 
& \left[ X_{2},X_{3}\right] =-2X_{1} \\ 
\left[ E,\left( iE\right) \right] =0 &  & \left[ H,E\right] =2E &  & \left[
H,\left( iE\right) \right] =2\left( iE\right) \\ 
\left[ X_{1},E\right] =-X_{3} &  & \left[ X_{1},\left( iE\right) \right] =H
&  & \left[ X_{1},H\right] =2X_{1}-4\left( iE\right) \\ 
\left[ X_{2},E\right] =H &  & \left[ X_{2},\left( iE\right) \right] =X_{3} & 
& \left[ X_{2},H\right] =2X_{2}-4E \\ 
\left[ X_{3},E\right] =2\left( iE\right) &  & \left[ X_{3},\left( iE\right) %
\right] =-2E &  & \left[ X_{3},H\right] =0%
\end{array}
\label{sl(2,C) Lie alg}
\end{equation}

The Killing form for $\mathfrak{sl}_{2}(\mathbb{C})$ is%
\begin{equation}
\kappa (X,Y):=~\mathsf{tr}\,{(ad}\left( {X}\right) {ad}\left( {Y}\right) {)}%
=4~\mathsf{tr}\,{(XY)},  \label{killing sl2C}
\end{equation}%
and the restriction to $\mathfrak{su}_{2}$ is negative definite. Hence we
take%
\begin{equation}
\mathrm{k}_{0}(X,Y)=-\dfrac{1}{4}\,\mathrm{Im}\,\kappa (X,Y)
\label{sym bil form on sl2}
\end{equation}%
that is a symmetric nondegenerate Ad-invariant bilinear form turning $%
\mathfrak{b}$\emph{\ }and\emph{\ }$\mathfrak{su}_{2}$\emph{\ }into\emph{\
isotropic subspaces}. It induces the linear bijection $\gamma :\mathfrak{su}%
_{2}\rightarrow \mathfrak{b}^{\ast }$ defined on the basis as 
\begin{equation}
\begin{array}{ccccc}
\gamma (X_{1})=-\mathbf{e} & , & \gamma (X_{2})=\mathbf{\tilde{e}} & , & 
\gamma (X_{3})=-2\mathbf{h}%
\end{array}
\label{psi su2->b*}
\end{equation}%
or its dual $\gamma ^{\ast }:\mathfrak{b}\rightarrow \mathfrak{su}_{2}^{\ast
}$ 
\begin{equation}
\begin{array}{ccccc}
\gamma ^{\ast }(E)=-\mathbf{x}_{1} & , & \gamma ^{\ast }(iE)=\mathbf{x}_{2}
& , & \gamma ^{\ast }(H)=-2\mathbf{x}_{3}%
\end{array}
\label{psi b->su2*}
\end{equation}

On $\mathfrak{su}_{2}$ there is the standard Killing form $\kappa $ defined
as 
\begin{equation*}
\kappa _{\mathfrak{su}_{2}}(X,Y):=4~\mathsf{tr}\,{(XY)},
\end{equation*}%
such that on the basis $\left\{ X_{1},X_{2},X_{3},\right\} $ it is $\left(
\kappa _{\mathfrak{su}_{2}}\right) _{ij}=-8\delta _{ij}$. Then, we may use
it to set a linear bijection $\zeta :\mathfrak{su}_{2}\longrightarrow 
\mathfrak{su}_{2}^{\ast }$ such that%
\begin{equation*}
\left\langle \zeta \left( X\right) ,Y\right\rangle =\kappa _{\mathfrak{su}%
_{2}}(X,Y)
\end{equation*}%
then, for the basis element we have 
\begin{equation*}
\left\langle \zeta \left( X_{i}\right) ,X_{j}\right\rangle =\kappa _{%
\mathfrak{su}_{2}}(X_{i},X_{j})=-8\delta _{ij}\Longrightarrow \mathbf{x}%
_{i}=-8\zeta \left( X_{i}\right)
\end{equation*}%
that implies explicitly%
\begin{equation}
\begin{array}{ccccc}
\zeta \left( X_{1}\right) =-\dfrac{1}{8}\mathbf{x}_{1} & , & \zeta \left(
X_{2}\right) =-\dfrac{1}{8}\mathbf{x}_{2} & , & \zeta \left( X_{3}\right) =-%
\dfrac{1}{8}\mathbf{x}_{3}%
\end{array}
\label{su2 to su2*}
\end{equation}

With these bijection we get the isomorphism $\vartheta :\gamma \circ \zeta
^{-1}\circ \gamma ^{\ast }:\mathfrak{b}\longrightarrow \mathfrak{b}^{\ast }$
such that%
\begin{equation}
\begin{array}{ccccc}
\vartheta \left( E\right) =-8\mathbf{e} & , & \vartheta \left( iE\right) =-8%
\mathbf{\tilde{e}} & , & \vartheta \left( H\right) =-32\mathbf{h}%
\end{array}
\label{b to b*}
\end{equation}%
and we define the linear bijection $\sigma :\mathfrak{sl}_{2}(\mathbb{C}%
)\longrightarrow \mathfrak{sl}_{2}(\mathbb{C})^{\ast }$ from the bijections $%
\left( \ref{su2 to su2*}\right) $ and $\left( \ref{b to b*}\right) $, so we
get 
\begin{equation}
\begin{array}{ccccc}
\sigma \left( X_{1}\right) =-\dfrac{1}{8}\mathbf{x}_{1} & , & \sigma \left(
X_{2}\right) =-\dfrac{1}{8}\mathbf{x}_{2} & , & \sigma \left( X_{3}\right) =-%
\dfrac{1}{8}\mathbf{x}_{3} \\ 
\sigma \left( E\right) =-8\mathbf{e} & , & \sigma \left( iE\right) =-8%
\mathbf{\tilde{e}} & , & \sigma \left( H\right) =-32\mathbf{h}%
\end{array}
\label{sigma def}
\end{equation}%
that is symmetric $\left\langle \sigma \left( X\right) ,Y\right\rangle
=\left\langle \sigma \left( Y\right) ,X\right\rangle $ because it is defined
from a symmetric bilinear form.

\subsection{Semidirect product of $SL(2,\mathbb{C})$ with $\mathfrak{sl}_{2}(%
\mathbb{C})$}

Then, we shall work on $H_{0}=SL(2,\mathbb{C})$ and its Lie algebra $%
\mathfrak{h}_{0}=\mathfrak{sl}_{2}(\mathbb{C})$ by considering the
semidirect product Lie group $H_{1}=SL(2,\mathbb{C})$ $\circledS \mathfrak{sl%
}_{2}(\mathbb{C})$ with the vector space $\mathfrak{sl}_{2}(\mathbb{C})$
regarded as the representation space for the \emph{adjoint action} in the
right action structure of semidirect product 
\begin{equation*}
\left( a,X\right) \bullet \left( b,Y\right) =\left( ab,\mathrm{Ad}%
_{b^{-1}}^{0}X+Y\right)
\end{equation*}

We assume $\mathfrak{sl}_{2}(\mathbb{C})$ is equipped with a nondegenerate
symmetric bilinear form $\left( \ref{sym bil form on sl2}\right) $ 
\begin{equation*}
\mathrm{k}_{0}(X,Y)=-\dfrac{1}{4}\,\mathrm{Im}\,\kappa (X,Y)=-\mathrm{Im}\,\mathsf{tr}\,{(XY)}
\end{equation*}

Let $h_{0}^{+},h_{0}^{-}$ be generic element of $SU\left( 2\right) $ and $B$%
, respectively, then 
\begin{equation*}
\begin{array}{ccc}
h_{0}^{+}=\left( 
\begin{array}{cc}
\alpha & \beta \\ 
-\bar{\beta} & \bar{\alpha}%
\end{array}%
\right) & , & h_{0}^{-}=\left( 
\begin{array}{cc}
a & b+ic \\ 
0 & a^{-1}%
\end{array}%
\right)%
\end{array}%
\end{equation*}%
with $\alpha ,\beta \in \mathbb{C}$, satisfying $\left\vert \alpha
\right\vert ^{2}+\left\vert \beta \right\vert ^{2}=1$, $a\in \mathbb{R}_{>0}$%
, and $b,c\in \mathbb{R}$. We have also an expression analogous to $\left( %
\ref{R2}\right) $ for the adjoint action of $\mathfrak{b}$ on $\mathfrak{su}%
_{2}$,%
\begin{equation*}
\mathrm{Ad}_{h_{-}^{0}}^{0}X_{+}^{0}=\left( h_{-}^{0}\right)
^{X_{+}^{0}}\left( h_{-}^{0}\right) ^{-1}+\gamma ^{-1}\left( Ad_{\left(
h_{-}^{0}\right) ^{-1}}^{\ast }\gamma \left( X_{+}^{0}\right) \right)
\end{equation*}%
and each term in the rhs can be calculated explicitly, for $%
X_{+}^{0}=x_{1}X_{1}+x_{2}X_{2}+x_{3}X_{3}$, to have%
\begin{eqnarray*}
\Pi _{-}\mathrm{Ad}_{h_{-}^{0}}^{0}X_{+}^{0} &=&\left( h_{-}^{0}\right)
^{Z_{+}^{0}}\left( h_{-}^{0}\right) ^{-1} \\
&=&\left( 2bcx_{1}+\left( b^{2}-c^{2}+a^{2}-\frac{1}{a^{2}}\right)
x_{2}+2acx_{3}\right) E \\
&&+\left( x_{1}\left( c^{2}-b^{2}+a^{2}-\frac{1}{a^{2}}\right)
+x_{2}2bc-x_{3}2ab\right) \left( iE\right) \\
&&-\left( x_{1}\dfrac{c}{a}+x_{2}\frac{b}{a}\right) H
\end{eqnarray*}%
\begin{eqnarray*}
\Pi _{+}\mathrm{Ad}_{h_{-}^{0}}^{0}X_{+}^{0} &=&\gamma ^{-1}\left(
Ad_{\left( h_{-}^{0}\right) ^{-1}}^{\ast }\gamma \left( X_{+}^{0}\right)
\right) \\
&=&x_{1}\dfrac{1}{a^{2}}X_{1}+x_{2}\frac{1}{a^{2}}X_{2}+\left( x_{1}\dfrac{b%
}{a}-x_{2}\frac{c}{a}+x_{3}\right) X_{3}
\end{eqnarray*}

In order to built the variables $\left( \mathbf{\tilde{\Omega}}_{0},\mathbf{N%
}_{0},\mathbf{\tilde{\Gamma}}_{0},\mathbf{M}_{0},\mathbf{R}_{0}^{-}\right) $
introduced in eq. $\left( \ref{Omega 0 - Gamma 0}\right) $, we shall need
the adjoint action of $B$ on $\mathfrak{sl}_{2}(\mathbb{C})$ such that for $%
X\in \mathfrak{sl}_{2}(\mathbb{C})$ we write%
\begin{equation*}
X=x_{1}X_{1}+x_{2}X_{2}+x_{3}X_{3}+x_{H}H+x_{E}E+x_{\left( iE\right) }\left(
iE\right)
\end{equation*}%
and then%
\begin{eqnarray*}
&&\mathrm{Ad}_{h_{-}^{0}}^{0}X \\
&=&x_{1}\dfrac{1}{a^{2}}X_{1}+x_{2}\frac{1}{a^{2}}X_{2}+\left( x_{1}\dfrac{b%
}{a}-x_{2}\frac{c}{a}+x_{3}\right) X_{3} \\
&&+\left( 2bcx_{1}+\left( b^{2}-c^{2}+a^{2}-\frac{1}{a^{2}}\right)
x_{2}+2acx_{3}+x_{E}a^{2}-2x_{H}ba\right) E \\
&&+\left( x_{1}\left( c^{2}-b^{2}+a^{2}-\frac{1}{a^{2}}\right)
+x_{2}2bc-x_{3}2ab+x_{\left( iE\right) }a^{2}-2x_{H}ca\right) \left(
iE\right) \\
&&-\left( x_{1}\dfrac{c}{a}+x_{2}\frac{b}{a}-x_{H}\right) H
\end{eqnarray*}%
The operators $\mathbb{A}_{\pm }^{0}\left( h_{-}\right) $ are%
\begin{equation}
\begin{array}{l}
\mathbb{A}_{+}^{0}\left( h_{-}^{0}\right) X=\Pi _{+}X+\dfrac{1}{a^{2}}\left(
\left( b^{2}+c^{2}+\dfrac{1}{a^{2}}-a^{2}\right) x_{2}-2acx_{3}\right) E \\ 
\qquad \qquad \qquad +\dfrac{1}{a^{2}}\left( \left( b^{2}+c^{2}+\dfrac{1}{%
a^{2}}-a^{2}\right) x_{1}+2abx_{3}\right) \left( iE\right) \\ 
\qquad \qquad \qquad +\dfrac{1}{a}\left( x_{1}c+x_{2}b\right) H%
\end{array}
\label{Ad + Ad}
\end{equation}%
and 
\begin{equation}
\begin{array}{l}
\mathbb{A}_{-}^{0}\left( h_{-}^{0}\right) X=\dfrac{1}{a^{2}}\left( \left(
a^{2}-\dfrac{1}{a^{2}}-b^{2}-c^{2}\right) x_{2}+2acx_{3}+a^{2}x_{E}\right) E
\\ 
\qquad \qquad \qquad +\dfrac{1}{a^{2}}\left( \left( a^{2}-\dfrac{1}{a^{2}}%
-b^{2}-c^{2}\right) x_{1}-2abx_{3}+a^{2}x_{\left( iE\right) }\right) \left(
iE\right) \\ 
\qquad \qquad \qquad -\left( \dfrac{c}{a}x_{1}+\dfrac{b}{a}%
x_{2}-x_{H}\right) H%
\end{array}
\label{Ad - Ad}
\end{equation}

Integrable systems are built in fiber on points $\left(
h_{-}^{0},Z_{-}^{0}\right) $ with $\Pi _{-}^{\ast }\sigma \left(
Z_{-}^{0}\right) $ being a character of $\mathfrak{b}$. Because the
coadjoint left action of $b\in B$ on $\mathfrak{b}^{\ast }$ is 
\begin{equation}
\begin{array}{lllll}
Ad_{\tilde{b}^{-1}}^{\ast }\mathbf{e}=2\dfrac{b}{a}\mathbf{h}+a^{-2}\mathbf{e%
} & , & Ad_{\tilde{b}^{-1}}^{\ast }\mathbf{\tilde{e}}=2\dfrac{c}{a}\mathbf{h}%
+a^{-2}\mathbf{\tilde{e}} & , & Ad_{\tilde{b}^{-1}}^{\ast }\mathbf{h}=%
\mathbf{h}%
\end{array}
\label{coadjoint act B on b*}
\end{equation}%
we conclude that the only character in $\mathfrak{b}$ is $0$, so $\Pi
_{-}^{\ast }\sigma \left( Z_{-}^{0}\right) =0$

\subsection{Dirac brackets on $\mathcal{N}_{2}\left( h_{1}^{-},0\right)
\subset SL(2,\mathbb{C})\circledS \mathfrak{sl}_{2}(\mathbb{C})$}

The last result of the previous subsections allows to write the Dirac
bracket on $\mathcal{N}_{2}\left( h_{1}^{-},Z_{1}^{-}\right) $, given in $%
\left( \ref{Dirac bracket G+xg+ I}\right) $, for $Z_{1}^{-}$ a character of $%
\mathfrak{b}$ as 
\begin{eqnarray}
&&\left\{ \mathcal{F},\mathcal{H}\right\} ^{D}\left( h,Z_{+}\right)
\label{Dirac brac SL2C} \\
&=&\left\langle h\mathbf{d}\mathcal{F},\mathbb{A}_{+}^{0}\left( h_{-}\right)
\Pi _{+}\sigma ^{-1}\left( \delta \mathcal{H}\right) \right\rangle
-\left\langle h\mathbf{d}\mathcal{H},\mathbb{A}_{+}^{0}\left( h_{-}\right)
\Pi _{+}\sigma ^{-1}\left( \delta \mathcal{F}\right) \right\rangle  \notag \\
&&-\left\langle \sigma \left( Z_{+}\right) ,\Pi _{+}[\mathbb{A}%
_{+}^{0}\left( h_{-}\right) \Pi _{+}\sigma ^{-1}\left( \delta \mathcal{F}%
\right) ,\mathbb{A}_{+}^{0}\left( h_{-}\right) \Pi _{+}\sigma ^{-1}\left(
\delta \mathcal{H}\right) ]\right\rangle  \notag
\end{eqnarray}%
The last term in the rhs can be written as 
\begin{eqnarray*}
&&\left\langle \sigma \left( Z_{+}\right) ,\Pi _{+}[\mathbb{A}_{+}^{0}\left(
h_{-}\right) \Pi _{+}\sigma ^{-1}\left( \delta \mathcal{F}\right) ,\mathbb{A}%
_{+}^{0}\left( h_{-}\right) \Pi _{+}\sigma ^{-1}\left( \delta \mathcal{H}%
\right) ]\right\rangle \\
&=&16\left( \delta \mathcal{F}\right) _{i}\left( \delta \mathcal{H}\right)
_{j}\varepsilon _{ijk}\left( z_{i}-\mathcal{B}_{k}\left( h_{-},Z_{+}\right)
\right)
\end{eqnarray*}%
where 
\begin{equation*}
\mathcal{B}\left( h_{-},Z_{+}\right) =\dfrac{b}{a}z_{3}X_{1}-\dfrac{c}{a}%
z_{3}X_{2}+\left( \dfrac{b}{a}z_{1}-\dfrac{c}{a}z_{2}-\left( \dfrac{b^{2}}{%
a^{2}}+\dfrac{c^{2}}{a^{2}}+\dfrac{1}{a^{4}}-1\right) z_{3}\right) X_{3}
\end{equation*}%
Observe that

\begin{eqnarray*}
\vec{\nabla}_{z}\times \mathcal{B}\left( h_{-},Z_{+}\right) &=&0 \\
&& \\
\vec{\nabla}_{z}\cdot \mathcal{B}\left( h_{-},Z_{+}\right) &=&-\left( \dfrac{%
b^{2}}{a^{2}}+\dfrac{c^{2}}{a^{2}}+\dfrac{1}{a^{4}}-1\right)
\end{eqnarray*}%
so it can be regarded as a static magnetic field associated to a magnetic
monopole charge density 
\begin{equation*}
\rho _{m}=-\dfrac{1}{a^{2}}\left( b^{2}+c^{2}+\dfrac{1}{a^{2}}-a^{2}\right) =%
\dfrac{1}{a^{2}}\mathrm{tr}\left( h_{-}Hh_{-}^{\dag }\right)
\end{equation*}%
For \emph{left invariant functions} the Poisson-Dirac bracket reduces to 
\begin{equation}
\left\{ \mathcal{F}^{L},\mathcal{H}^{L}\right\} ^{D}\left( h,Z\right)
=-16\left( \delta \mathcal{F}\right) _{i}\left( \delta \mathcal{H}\right)
_{j}\varepsilon _{ijk}\left( z_{i}-\mathcal{B}_{k}\left( h_{-},Z_{+}\right)
\right)  \label{Dirac brack for left inv fun}
\end{equation}

Now we shall consider the semidirect product Lie group $H_{2}=H_{1}\circledS 
\mathfrak{h}_{1}$, where $H_{1}=SL(2,\mathbb{C})\circledS \mathfrak{sl}_{2}(%
\mathbb{C})$. As in previous sections, in both cases we use the right action
structure of semidirect product $\left( \ref{Hxh-1}\right) $.We define on $%
\mathfrak{h}_{1}$ the nondegenerate symmetric bilinear form%
\begin{equation*}
\mathrm{k}_{1}\left( (X,U),\left( Y,V\right) \right) =\frac{1}{2}\left( 
\mathrm{k}_{0}(X,V)+\mathrm{k}_{0}(U,Y)\right)
\end{equation*}

Regarding $H_{2}=H_{1}\circledS \mathfrak{h}_{1}$ as a phase space, where $%
H_{1}$ inherits the factorization of $SL(2,\mathbb{C)}=H_{0}^{+}\times
H_{0}^{-}$ being $H_{0}^{+}=SU\left( 2\right) $ and $H_{0}^{-}=B$, the group
of $2\times 2$ complex upper triangular matrices with positive real diagonal
elements and determinant equal 1. This means that $H_{2}=H_{1}^{+}\times
H_{1}^{-}$ where $H_{1}^{\pm }=H_{0}^{\pm }\circledS \mathfrak{h}_{0}^{\pm }$%
.

In this framework, we shall study dynamical system on the fibration $H_{2}%
\overset{\Psi _{2}}{\longrightarrow }H_{2}^{-}$ by applying the Dirac
procedure explained above. We use the results obtained in section $\left( %
\ref{nested}\right) $, describing in terms of the variables $\left( \mathbf{%
\Omega }_{1},\mathbf{\Gamma }_{1}\right) $, see eq. $\left( \ref{Omega-Gamma}%
\right) $, a collective system on $\mathcal{N}\left(
h_{1}^{-},Z_{1}^{-}\right) $ with equations of motion $\left( \ref{Kirchoff
nested eqs 1}\right) $ $\left( \ref{collective ham eq. on N}\right) $%
\begin{equation*}
\left\{ 
\begin{array}{l}
\left( h_{1}^{+}\right) ^{-1}\dot{h}_{1}^{+}=\mathbf{\Omega }_{1}^{+} \\ 
\\ 
\mathbf{\dot{\Gamma}}_{1}=\Pi _{-}[\mathbf{\Gamma }_{1},\mathbf{\Omega }%
_{1}^{+}]%
\end{array}%
\right.
\end{equation*}

Regarding $H_{1}$ as the semidirect product $H_{1}=H_{0}\circledS \mathfrak{h%
}_{0}$, we introduce the variables $\left( \mathbf{\tilde{\Omega}}_{0},%
\mathbf{N}_{0},\mathbf{\tilde{\Gamma}}_{0},\mathbf{M}_{0},\mathbf{R}%
_{0}^{-}\right) $ as defined in eq. $\left( \ref{Omega 0 - Gamma 0}\right) $%
, yielding the set of differential equations%
\begin{equation}
\left\{ 
\begin{array}{l}
\left( h_{0}^{+}\right) ^{-1}\dfrac{d}{dt}h_{0}^{+}=\mathbf{\tilde{\Omega}}%
_{0}^{+} \\ 
\\ 
\dfrac{d}{dt}Z_{0}^{+}=-\left[ \mathbf{\tilde{\Omega}}_{0}^{+},Z_{0}^{+}%
\right] +\mathbf{N}_{0} \\ 
\\ 
\dfrac{d}{dt}\mathbf{\tilde{\Gamma}}_{0}=\Pi _{-}\left[ \mathbf{\tilde{\Gamma%
}}_{0},\mathbf{\tilde{\Omega}}_{0}^{+}\right] \\ 
\\ 
\dfrac{d}{dt}\mathbf{M}_{0}=\Pi _{-}\left[ \mathbf{\tilde{\Gamma}}_{0},%
\mathbf{N}_{0}^{+}\right] +\Pi _{-}\left[ \mathbf{M}_{0},\mathbf{\tilde{%
\Omega}}_{0}^{+}\right]%
\end{array}%
\right.  \label{SL2C eqs}
\end{equation}%
since $Z_{0}^{-}=0$ and $\mathbf{R}_{0}^{-}=0$.

\subsection{Solving by factorization}

In particular, we consider a collective hamiltonian on $H_{2}$ 
\begin{equation*}
\mathcal{H}^{\left( 2\right) }(h_{1},Z_{1}):=\mathsf{h}^{\left( 2\right)
}\left( \Phi _{B}^{2}(h_{1},Z_{1})\right)
\end{equation*}%
with 
\begin{equation*}
\Phi _{B}^{\left( 2\right) }(h_{1},Z_{1})=\gamma \left( \mathrm{Ad}%
_{h_{1}}^{1}\gamma ^{-1}\left( \sigma \left( Z_{1}\right) \right) \right) =%
\mathrm{Ad}_{h_{1}^{-1}}^{1\ast }\sigma \left( Z_{1}\right)
\end{equation*}%
and the Hamilton function on $\mathfrak{h}_{1}^{\ast }$ as $\mathsf{h}%
^{\left( 2\right) }:\mathfrak{h}_{1}^{\ast }\longrightarrow \mathbb{R}$ 
\begin{equation*}
\mathsf{h}^{\left( 2\right) }\left( X_{0},Y_{0}\right) =-\frac{1}{16}\,\mathrm{Re}\,\kappa \left( X_{0},Y_{0}\right)
\end{equation*}%
which is $\mathrm{Ad}^{1}$ since it is naturally $\mathrm{Ad}^{0}$
invariant. The Killing form on $\kappa :\mathfrak{sl}_{2}\mathbb{C\times 
\mathfrak{sl}}_{2}\mathbb{\mathbb{C}\longrightarrow C}$ is defined in eq. $%
\left( \ref{killing sl2C}\right) $.

Its Legendre transform, $\mathcal{L}_{\mathsf{h}^{\left( 2\right) }}:%
\mathfrak{sl}_{2}\mathbb{C\oplus }\mathfrak{sl}_{2}\mathbb{C}\longrightarrow 
\mathfrak{sl}_{2}^{\ast }\mathbb{C\oplus }\mathfrak{sl}_{2}^{\ast }\mathbb{C}
$, is%
\begin{equation*}
\mathcal{L}_{\mathsf{h}^{\left( 2\right) }}\left( X_{0},Y_{0}\right) =\frac{i%
}{2}\left( X_{0},Y_{0}\right)
\end{equation*}

Remembering that $\sigma \left( Z_{1}^{-}\right) $ is a character of $%
\mathfrak{h}_{1}^{-}$, we conclude that $\left( \sigma \left(
X_{0}^{-}\right) ,\sigma \left( Y_{0}^{-}\right) \right) \in \mathrm{ch~}%
\mathfrak{h}_{0}^{-}\oplus \mathrm{ch~}\mathfrak{h}_{0}^{-}$, and we saw in $%
\left( \ref{coadjoint act B on b*}\right) $ that in $\mathfrak{b}$ there are
no nontrivial characters, so $X_{0}^{-}=Y_{0}^{-}=0$. Therefore%
\begin{equation*}
\mathcal{L}_{\mathsf{h}^{\left( 2\right) }}\left( X_{0},Y_{0}\right) =\frac{i%
}{2}\left( X_{0}^{+},Y_{0}^{+}\right)
\end{equation*}

The AKS scheme states that the $H_{1}^{\pm }$ factors of the exponential
curve 
\begin{equation}
k\left( t\right) =e^{t\mathcal{L}_{\mathsf{h}^{\left( 2\right) }}\left(
X_{0},Y_{0}\right) }=e^{\frac{i}{2}t\left( X_{0}^{+},Y_{0}^{+}\right) }
\label{sol exp 0}
\end{equation}%
solve the original system of differential equations, for some time
independent $\left( X_{0}^{+},Y_{0}^{+}\right) \in \mathfrak{h}_{1}^{+}$. In
order to analyze what is happening at the level of $H_{0}$ and $\mathfrak{h}%
_{0}$, we write these exponential by using the definition of the
exponential, eq. $\left( \ref{Hxh-4}\right) $,%
\begin{equation}
e^{\frac{i}{2}t\left( X_{0}^{+},Y_{0}^{+}\right) }=\left( e^{i\frac{t}{2}%
X_{0}^{+}},-\sum_{n=1}^{\infty }\frac{\left( -1\right) ^{n}}{n!}\left( \frac{%
i}{2}t\right) ^{n}\left( \mathrm{ad}_{X_{0}^{+}}^{1}\right)
^{n-1}Y_{0}^{+}\right)  \label{sol exp 1}
\end{equation}

Using the result $\left( \ref{factorization semidirect 2}\right) $, we have 
\begin{equation}
\left\{ 
\begin{array}{l}
\Pi _{+}e^{\frac{i}{2}t\left( X_{0}^{+},Y_{0}^{+}\right) }=\left(
h_{0}^{+}\left( t\right) ,-\mathrm{Ad}_{k_{-}^{0}}^{0}\mathbb{A}%
_{0}^{+}\left( k_{0}^{-}\right) \sum\limits_{n=1}^{\infty }\dfrac{\left(
-1\right) ^{n}}{n!}\left( \dfrac{i}{2}t\right) ^{n}\left( \mathrm{ad}%
_{X_{0}^{+}}^{1}\right) ^{n-1}Y_{0}^{+}\right) \\ 
\\ 
\Pi _{-}e^{\frac{i}{2}t\left( X_{0}^{+},Y_{0}^{+}\right) }=\left(
k_{0}^{-}\left( t\right) ,-\Pi _{-}\mathbb{A}_{0}^{-}\left( k_{0}^{-}\right)
\sum\limits_{n=1}^{\infty }\dfrac{\left( -1\right) ^{n}}{n!}\left( \dfrac{i}{%
2}t\right) ^{n}\left( \mathrm{ad}_{X_{0}^{+}}^{1}\right)
^{n-1}Y_{0}^{+}\right)%
\end{array}%
\right.  \notag
\end{equation}%
where we wrote $e^{i\frac{t}{2}X_{0}^{+}}=h_{0}^{+}\left( t\right)
k_{0}^{-}\left( t\right) $.

The exponential for $X_{0}^{+}\in \mathfrak{su}_{2}$ can be explicitly
computed from the relations 
\begin{equation*}
\left\{ 
\begin{array}{l}
\left( X_{0}^{+}\right) ^{2n}=\left( -1\right) ^{n}\left\vert \left\vert
X_{0}^{+}\right\vert \right\vert ^{n}I \\ 
\\ 
\left( X_{0}^{+}\right) ^{2n+1}=\left( -1\right) ^{n}\left\vert \left\vert
X_{0}^{+}\right\vert \right\vert ^{n}X_{0}^{+}%
\end{array}%
\right.
\end{equation*}%
where $\left\vert \left\vert X\right\vert \right\vert =\sqrt{\mathrm{det}X}$%
, then 
\begin{equation*}
\exp \left( i\frac{t}{2}X_{0}^{+}\right) =\cosh {\left( t\frac{\left\vert
\left\vert X_{0}^{+}\right\vert \right\vert }{2}\right) }+iX_{0}^{+}\sinh {%
\left( t\frac{\left\vert \left\vert X_{0}^{+}\right\vert \right\vert }{2}%
\right) }
\end{equation*}%
Without loss of generality, we may take $\left\vert \left\vert X\right\vert
\right\vert =1$, so%
\begin{equation*}
\exp \left( i\frac{t}{2}X_{0}^{+}\right) =\cosh {\left( \frac{t}{2}\right) }%
+iX_{0}^{+}\sinh {\left( \frac{t}{2}\right) }
\end{equation*}

In order to obtain the factors $h_{0}^{+}\left( t\right) $ and $%
k_{0}^{-}\left( t\right) $, we write the time independent vector $%
X_{0}^{+}=a_{1}X_{1}+a_{2}X_{2}+a_{3}X_{3}$, so $\left\vert \left\vert
X\right\vert \right\vert =1$ is equivalent to $%
a_{1}^{2}+a_{2}^{2}+a_{3}^{2}=1$. Thus, the curve in $SL(2,\mathbb{C)}$%
\begin{equation*}
\exp \left( i\frac{t}{2}X_{0}^{+}\right) =\left( 
\begin{array}{cc}
\cosh \left( t/2\right) -a_{3}\sinh \left( t/2\right) & -\left(
a_{1}-ia_{2}\right) \sinh \left( t/2\right) \\ 
-\left( a_{1}+ia_{2}\right) \sinh \left( t/2\right) & \cosh \left(
t/2\right) +a_{3}\sinh \left( t/2\right)%
\end{array}%
\right)
\end{equation*}%
can be factorized as $\exp \left( i\frac{t}{2}X_{0}^{+}\right)
=h_{0}^{+}\left( t\right) k_{0}^{-}\left( t\right) $, for $h_{0}^{+}\left(
t\right) \subset SU(2)$ and $k_{0}^{-}\left( t\right) \subset B$, with%
\begin{align}
h_{0}^{+}\left( t\right) & =\left( 
\begin{array}{ccc}
\dfrac{\cosh {\left( t/2\right) }-a_{3}\sinh {\left( t/2\right) }}{\sqrt{%
\cosh {t}-a_{3}\sinh {t}}} &  & \dfrac{\left( a_{1}-ia_{2}\right) \sinh {%
\left( t/2\right) }}{\sqrt{\cosh {t}-a_{3}\sinh {t}}} \\ 
&  &  \\ 
-\dfrac{\left( a_{1}+ia_{2}\right) \sinh {\left( t/2\right) }}{\sqrt{\cosh {t%
}-a_{3}\sinh {t}}} &  & \dfrac{\cosh {\left( t/2\right) }-a_{3}\sinh {\left(
t/2\right) }}{\sqrt{\cosh {t}-a_{3}\sinh {t}}}%
\end{array}%
\right)  \label{SU(2) factor} \\
&  \notag \\
k_{0}^{-}\left( t\right) & =\left( 
\begin{array}{ccc}
\sqrt{\cosh {t}-a_{3}\sinh {t}} &  & \dfrac{-\left( a_{1}-ia_{2}\right)
\sinh {t}}{\sqrt{\cosh {t}-a_{3}\sinh {t}}} \\ 
&  &  \\ 
0 &  & \left( \sqrt{\cosh {t}-a_{3}\sinh {t}}\right) ^{-1}%
\end{array}%
\right)  \label{B factor}
\end{align}

Let us now address the second component in the exponential solution $\left( %
\ref{sol exp 1}\right) $. In order to handle it more easily, we shall
introduce some notation. For the basis $\left\{ X_{1},X_{2},X_{3}\right\} $
on $\mathfrak{su}\left( 2\right) $, eq. $\left( \ref{su2 basis}\right) $, we
introduce the following notation: we write each element $X$ in $\mathfrak{su}%
\left( 2\right) $ as 
\begin{equation*}
X=\mathbf{x}\cdot \mathbf{X}:=x_{1}X_{1}+x_{2}X_{2}+x_{3}X_{3}
\end{equation*}%
where $\mathbf{x}:=\left( x_{1},x_{2},x_{3}\right) $ is the real $3$-vector
formed by the components of $X$. Also, the relations 
\begin{eqnarray*}
\left[ X_{i},X_{j}\right] &=&-2\epsilon _{ijk}X_{k} \\
&& \\
X_{i}X_{j} &=&-\delta _{ij}I-\epsilon _{ijk}X_{k}
\end{eqnarray*}%
will be useful. With the above notation, the adjoint Lie bracket can be
written as 
\begin{equation*}
\mathrm{ad}_{X}Y=-2\left( \mathbf{x}\times \mathbf{y}\right) \cdot \mathbf{X}
\end{equation*}%
where $Y:=\mathbf{y}\cdot \mathbf{X}$, and $"\times "$ means the standard
vector product in $\mathbb{R}^{3}$; and 
\begin{equation*}
\left( \mathrm{ad}_{X}\right) ^{n}Y=\left( -2\right) ^{n}\left[ \underset{n%
\text{ times}}{\underbrace{\mathbf{x}\times (\cdots (\mathbf{x}\times }}%
\mathbf{y})\cdots )\right] \cdot \mathbf{X}
\end{equation*}

The next Lemma and its Corollary will help us to write the exponential terms
in $e^{\frac{i}{2}t\left( X_{+}^{0},Y_{+}^{0}\right) }$ in a more friendly
way.

\begin{description}
\item[Lemma:] \textit{Let} $\mathbf{x},\mathbf{y}\in \mathbb{R}^{3}$ \textit{%
be a pair of vectors. Then, for arbitrary }$n\in \mathbb{N}$\textit{,}%
\begin{equation*}
\underset{n\text{ times}}{\underbrace{\mathbf{x}\times (\cdots (\mathbf{x}%
\times }}\mathbf{y})\cdots )=\left\{ 
\begin{array}{lll}
\left( -1\right) ^{\frac{n}{2}+1}\left\Vert \mathbf{x}\right\Vert ^{n-2}%
\left[ \left( \mathbf{x}\cdot \mathbf{y}\right) \mathbf{x}-\left\Vert 
\mathbf{x}\right\Vert ^{2}\mathbf{y}\right] &  & \text{for }n\text{ even }
\\ 
&  &  \\ 
\left( -1\right) ^{\frac{n-1}{2}}\left\Vert \mathbf{x}\right\Vert ^{n-1}%
\mathbf{x}\times \mathbf{y} &  & \text{for }n\text{ odd}%
\end{array}%
\right.
\end{equation*}
\end{description}

Let us assume additionally that the $3$-vector $\mathbf{x}:=\left(
x^{1},x^{2},x^{3}\right) $ has unit norm $\mathbf{x}\cdot \mathbf{x}=1$,
then we can establish the following.

\begin{description}
\item[Corollary:] \textit{The map} $\left( \mathrm{ad}_{X_{0}^{+}}\right)
^{n}:\mathfrak{su}\left( 2\right) \rightarrow \mathfrak{su}\left( 2\right) $%
\textit{, for arbitrary }$n\in \mathbb{N}$\textit{,} \textit{has the formula}
\begin{equation*}
\left( \mathrm{ad}_{X_{0}^{+}}\right) ^{n}\left( Y_{0}^{+}\right) =\left\{ 
\begin{array}{lll}
\left( -1\right) ^{\frac{n}{2}+1}2^{n}\left[ \left( \mathbf{x}\cdot \mathbf{y%
}\right) X_{0}^{+}-Y_{0}^{+}\right] &  & \text{for }n\text{ even } \\ 
&  &  \\ 
\left( -1\right) ^{\frac{n-1}{2}}2^{n-1}\left[ X_{0}^{+},Y_{0}^{+}\right] & 
& \text{for }n\text{ odd}%
\end{array}%
\right.
\end{equation*}%
\textit{where} $Y_{0}^{+}:=\mathbf{y}\cdot \mathbf{X}$.
\end{description}

Thus, the second component in the rhs of eq. $\left( \ref{sol exp 1}\right) $%
\begin{eqnarray*}
&&-\left[ \sum_{n=1}^{\infty }\frac{\left( -1\right) ^{n}}{n!}\left( \frac{it%
}{2}\right) ^{n}\left( \mathrm{ad}_{X_{0}^{+}}\right) ^{n-1}\right] Y_{0}^{+}
\\
&=&\frac{i}{2}\left( t-\sinh {t}\right) \left( \mathbf{x}\cdot \mathbf{y}%
\right) X_{0}^{+}+\frac{i}{2}\left( \sinh {t}\right) Y_{0}^{+}+\frac{1}{4}%
\left( \cosh {t}-1\right) \left[ X_{0}^{+},Y_{0}^{+}\right]
\end{eqnarray*}%
and, finally, the factorization of this term is obtained by following the
relations $\left( \ref{factorization semidirect 2}\right) $, to get%
\begin{eqnarray}
X_{1}^{+}\left( t\right) &=&\Pi _{+}\mathrm{Ad}_{k_{0}^{-}\left( t\right)
}^{0}\mathbb{A}_{0}^{+}\left( k_{0}^{-}\left( t\right) \right) \left( \frac{i%
}{2}\left( t-\sinh {t}\right) \left( \mathbf{x}\cdot \mathbf{y}\right)
X_{0}^{+}\right.  \label{factorization sl2 +} \\
&&\qquad \qquad \qquad \left. +\frac{i}{2}\left( \sinh {t}\right) Y_{0}^{+}+%
\frac{1}{4}\left( \cosh {t}-1\right) \left[ X_{0}^{+},Y_{0}^{+}\right]
\right)  \notag
\end{eqnarray}%
and%
\begin{eqnarray}
X_{1}^{-}\left( t\right) &=&\mathbb{A}_{0}^{-}\left( k_{0}^{-}\left(
t\right) \right) \left( \frac{i}{2}\left( t-\sinh {t}\right) \left( \mathbf{x%
}\cdot \mathbf{y}\right) X_{0}^{+}\right.  \label{factorization sl2 -} \\
&&\qquad \qquad \left. +\frac{i}{2}\left( \sinh {t}\right) Y_{0}^{+}+\frac{1%
}{4}\left( \cosh {t}-1\right) \left[ X_{0}^{+},Y_{0}^{+}\right] \right) 
\notag
\end{eqnarray}

Using the expression analogous to $\left( \ref{R2}\right) $ for the adjoint
action of $\mathfrak{b}$ on $\mathfrak{su}_{2}$,%
\begin{equation*}
\mathrm{Ad}_{k_{-}^{0}}^{0}Z_{0}^{+}=\left( k_{0}^{-}\right)
^{Z_{0}^{+}}\left( k_{0}^{-}\right) ^{-1}+\gamma ^{-1}\left( Ad_{\left(
k_{0}^{-}\right) ^{-1}}^{\ast }\gamma \left( Z_{0}^{+}\right) \right)
\end{equation*}%
we have that, for $Z_{0}^{+}\in \mathfrak{su}_{2}$, 
\begin{equation*}
\left\{ 
\begin{array}{l}
\Pi _{+}\mathrm{Ad}_{k_{0}^{-}\left( t\right) }^{0}\mathbb{A}_{0}^{+}\left(
k_{0}^{-}\left( t\right) \right) Z_{0}^{+}=\gamma ^{-1}\left( Ad_{\left(
k_{0}^{-}\left( t\right) \right) ^{-1}}^{\ast }\gamma \left(
Z_{0}^{+}\right) \right) \\ 
\\ 
\mathbb{A}_{0}^{-}\left( k_{0}^{-}\left( t\right) \right) Z_{0}^{+}=\left(
k_{0}^{-}\right) ^{-1}\left( k_{0}^{-}\right) ^{Z_{0}^{+}}%
\end{array}%
\right.
\end{equation*}

In summary, joining the explicit forms of curves $\left( \ref{SU(2) factor},%
\ref{B factor}\right) $ on the group factors with the curves $\left( \ref%
{factorization sl2 +},\ref{factorization sl2 -}\right) $ we get the solution
for the hamiltonian system $\left( \ref{SL2C eqs}\right) $ with the Hamilton
function $\mathcal{H}^{\left( 2\right) }(h_{1},Z_{1})$ defined at the
beginning of this subsection.

\section{Conclusions}

Starting from a Poisson-Lie group, we have constructed integrable systems on
the semidirect product with its Lie algebra showing that most of the
standard issues of integrability and factorization are well suited in this
framework, allowing for wider class of systems. We have obtained the
Poisson-Lie structure on the semidirect product having a very simple
relation with the original one, although it does not map coboundaries into
coboundaries. This construction allows to supply each tangent bundle with
two Poisson structures: a nondegenerate one which is derived from the
canonical symplectic structure on the associated cotangent bundle through
some linear bijection, and a Poisson-Lie one inherited from the semidirect
product Lie group structure on the trivialization of tangent bundle.
Moreover, the construction can be iterated on iterated semidirect products
giving rise to a chain of phase spaces sharing both the Poisson structures.

On this chain, we get collective systems from the left translation momentum
map on the whole Lie group, and by means of the Dirac brackets we project
the collective dynamics onto a class of phase spaces which are isomorphic to
the tangent bundle of one of the factors. That means, Dirac bracket produces
non trivial integrable system on these phase subspaces from the collective
hamiltonians on the whole phase space. The associated hamiltonian vector
fields turn to be dressing vectors associated with the Poisson-Lie
structures of the previous step in the chain, and they fail in to be
hamiltonian in relation with the Poisson-Lie structure of the corresponding
level.

The system thus obtained are integrable by factorization of the Lie group in
the previous step, and it can be traced back to the factorization of the
initial Lie group of the chain. Moreover, for some special hamiltonians we
get a tower of integrable system where the dynamical system at each level
replicates one level down by projecting on the second component of the
semidirect product.

Thus, the presented construction provides a setting for new class of
integrable by factorization systems which are obtained from semidirect
products of Lie groups.

\section{Acknowledgments}

The authors thank to CONICET (Argentina) for financial support. S. Capriotti
wants to acknowledge financial support from IRSES project GEOMECH (nr.
246981) within the 7th European Community Framework Programme.


\end{document}